\documentclass{article}
\usepackage{bm,latexsym,amsmath,amssymb,amsfonts,fancyhdr,color,graphicx,multirow,slashed,cite}
\usepackage[a4paper,bottom=3cm,top=2.5cm,head=0mm,width=17cm,dvipdfm]{geometry}
\usepackage[usenames,dvipsnames,svgnames,table]{xcolor}
\usepackage[colorlinks=true,
            linkcolor=blue,
            urlcolor=blue,
            citecolor=green,          
            bookmarks=true,
            bookmarksnumbered=true,
            breaklinks=true,
            pdfstartview=FitBH
            ]{hyperref}
\usepackage[dotinlabels]{titletoc}
\usepackage{titlesec}
\usepackage{authblk,ulem}

\numberwithin{equation}{section}
\allowdisplaybreaks[4]

\titlelabel{\thetitle.\quad \hspace{-0.8em}}
\titlecontents{section}
              [1.5em]
              {\vspace{4mm} \large \boldsymbol}
              {\contentslabel{1em}}
              {\hspace*{-1em}}
              {\titlerule*[.5pc]{.}\contentspage}
\titlecontents{subsection}
              [3.5em]
              {\vspace{2mm}}
              {\contentslabel{1.8em}}
              {\hspace*{.3em}}
              {\titlerule*[.5pc]{.}\contentspage}
\titlecontents{subsubsection}
              [5.5em]
              {\vspace{2mm}}
              {\contentslabel{2.5em}}
              {\hspace*{.3em}}
              {\titlerule*[.5pc]{.}\contentspage}

\newcommand{\titledef}{Disentangle Neutrino Electromagnetic Properties with Atomic Radiative Pair Emission} % Insert Title here!!!
\hypersetup{ pdfauthor = {Shao-Feng Ge},
	     pdftitle = {\titledef}, % Insert title here!!!
	     pdfsubject = {}, % Insert subject here!!!
             pdfkeywords = {}, % Insert keywords here!!!
	     pdfcreator = {LaTeX with hyperref package},
	     pdfproducer = {dvips + ps2pdf} }

\definecolor{gesfblack}{rgb}{0,0,0}

\definecolor{gesfblue}{rgb}{0.08,0.42,0.76}

\definecolor{gesfgreen}{rgb}{0,1,0}

\definecolor{gesfgrey}{rgb}{0.5,0.5,0.5}

\definecolor{gesflanse}{rgb}{0.00,0.50,0.50}

\definecolor{gesfpurple}{rgb}{0.47,0.19,0.42}

\definecolor{gesfred}{rgb}{1,0,0}

\definecolor{gesfwhite}{rgb}{1,1,1}

\definecolor{gesfyellow}{rgb}{0.7,0.4,0.3}

\makeatletter
\newcommand{\ostar}{\mathbin{\mathpalette\make@circled\star}}
\newcommand{\make@circled}[2]{%
  \ooalign{$\m@th#1\smallbigcirc{#1}$\cr\hidewidth$\m@th#1#2$\hidewidth\cr}%
}
\newcommand{\smallbigcirc}[1]{%
  \vcenter{\hbox{\scalebox{0.8}{$\m@th#1\bigcirc$}}}%
}
\makeatother

\newcommand{\gsec}[1]{{\hypersetup{linkcolor=red}Sec.\,\ref{#1}\hypersetup{linkcolor=blue}}}
\newcommand{\gapp}[1]{{\hypersetup{linkcolor=red}Appendix \ref{#1}\hypersetup{linkcolor=blue}}}
\newcommand{\geqn}[1]{\hypersetup{linkcolor=blue}Eq.\,(\ref{#1})\hypersetup{linkcolor=blue}}
\newcommand{\gfig}[1]{{\hypersetup{linkcolor=violet}Fig.\,\ref{#1}\hypersetup{linkcolor=blue}}}
\newcommand{\gtab}[1]{{\hypersetup{linkcolor=gesflanse}Table~\ref{#1}\hypersetup{linkcolor=blue}}}

\definecolor{Orange}{cmyk}{0,0.61,0.87,0}
\definecolor{JungleGreen}{cmyk}{0.99,0,0.52,0}
\definecolor{OliveGreen}{cmyk}{0.64,0,0.95,0.40}
\definecolor{Brown}{cmyk}{0,0.81,1,0.60}
\definecolor{RoyalBlue}{cmyk}{0.71,0.53,0,0.12}
\definecolor{Gray}{cmyk}{0,0,0,0.40}
\definecolor{LightPink}{cmyk}{0.0,0.25,0,0}
\definecolor{LLightPink}{cmyk}{0.0,0.10,0,0}
\definecolor{LightBlue}{cmyk}{0.25,0,0,0}
\definecolor{LightGray}{cmyk}{0,0,0,0.2}

\usepackage{float}

\graphicspath{{figs/}}

\begin{document}
\fontsize{12pt}{14pt}\selectfont

\title{%\begin{flushright}
       %\mbox{\normalsize IPMU18-xxxx}
       %\end{flushright}
			 %\vskip 20pt
       \textbf{\fontsize{18pt}{20pt}\selectfont \titledef}} % Insert title here!!!
\author[1,2]{{\large Shao-Feng Ge} \footnote{\href{mailto:gesf@sjtu.edu.cn}{gesf@sjtu.edu.cn}}}
\affil[1]{Tsung-Dao Lee Institute \& School of Physics and Astronomy, Shanghai Jiao Tong University, Shanghai 200240, China}
\affil[2]{Key Laboratory for Particle Astrophysics and Cosmology (MOE) \& Shanghai Key Laboratory for Particle Physics and Cosmology, Shanghai Jiao Tong University, Shanghai 200240, China}
\author[3]{{\large Pedro Pasquini} \footnote{\href{mailto:pedrosimpas@g.ecc.u-tokyo.ac.jp}{pedrosimpas@g.ecc.u-tokyo.ac.jp}}}
\affil[3]{Department of Physics, University of Tokyo, Bunkyo-ku, Tokyo 113-0033, Japan}

\date{\today}

\maketitle

\begin{abstract}
\fontsize{12pt}{14pt}\selectfont
We elaborate the possibility of using the atomic radiative
emission of neutrino pair (RENP) to probe the neutrino
electromagnetic properties, including magnetic and electric
dipole moments, charge radius, and anapole. With the typical
$\mathcal O$(eV) momentum transfer, the atomic RENP is sensitive
to not just the tiny neutrino masses but also very light 
mediators to which the massless photon belongs.
The neutrino EM properties introduce
extra contribution besides the SM one induced by the
heavy $W/Z$ gauge bosons. Since the associated photon spectrum
is divided into several sections whose boundaries are determined
by the final-state neutrino masses, it is possible to
identify the individual neutrino EM form factor elements.
Most importantly, scanning the photon spectrum inside the
particular section with deviation from the SM prediction
once observed allows identification of the neutrino EM
form factor type. The RENP provides an ultimate way of
disentangling the neutrino EM properties to go beyond the
current experimental searches or observations.
\end{abstract}

\newpage

\section{Introduction}

The existence of neutrinos is indicated by the continuous
beta decay spectrum and informally proposed by Pauli in
his famous letter to the participants of the Tübingen
conference on radioactivity on December 4, 1930
\cite{Pauli:1991owm}. In exactly the same letter,
Pauli suggested that ``{\it the neutrino at rest is a
magnetic dipole with a certain moment}'' as the possible
interaction that neutrino may have. So, the neutrino EM
properties are actually the very first studied.
While Pauli only mentioned magnetic moment, the general
neutrino ($\nu$) coupling to 
photons ($A$) contain four independent form factor types 
at the vanishing momentum transfer limit ($q^2 \rightarrow 0)$
\cite{Nieves:1981zt,Kayser:1982br},
\begin{align}
  H_{EM}
=
  \overline \nu 
  \left[
-i \sigma_{\mu\nu} q^\nu 
  \left(
  \mu_\nu 
+ i \epsilon_\nu \gamma_5
  \right)
+ \left(
    \gamma_\mu - \frac{q_\mu \slashed q}{q^2}
  \right)q^2
  \left(
    \frac{\langle r^2_\nu\rangle}{6} 
  + a_\nu \gamma_5
  \right)
  \right]
  \nu 
  A^\mu(q),
\end{align}
where $\mu_\nu$, $\epsilon_\nu$, $\langle 
r^2_\nu\rangle$, and $a_\nu$ are the 
neutrino magnetic dipole moment (MDM), electric dipole
moment (EDM), charge 
radius, and anapole, respectively. Within the
three-neutrino paradigm, all the neutrino EM properties 
are Hermitian matrices. In the Standard Model (SM)
of particle physics, 
the tree-level coupling between neutrino and 
photon is zero \cite{Workman:2022ynf} while the 
loop-induced electromagnetic couplings are highly suppressed. 
While the neutrino charge radius and anapole are suppressed 
by the Fermi constant 
\cite{Dvornikov:2003js,Bernabeu:2000hf},
the neutrino magnetic moment and electric 
dipole are further suppressed by the neutrino mass
\cite{Fujikawa:1980yx,Pal:1981rm,Shrock:1982sc,
Dvornikov:2003js}.

Since the SM values for the neutrino EM properties are small,
their observation is a window 
to possible new physics beyond the Standard Model (BSM) 
\cite{Studenikin:2008bd,Novales-Sanchez:2008lyf,Brdar:2021xll}.
There are various ways of probing the neutrino EM
properties. The neutrino scattering experiments
measure the electron/nucleus recoil spectrum with
existing constraints ranging from $10^{-7}\mu_B$
($10^{-29}$\,cm$^2$) to $10^{-11}\mu_B$ ($10^{-32}$\,cm$^2$)
for neutrino magnetic/electric moments (charge
radius/anapole). Different neutrino sources have
characteristic energy scales, such as accelerators
($0.2\,{\rm GeV}  < E_\nu  < 20$\,GeV)
\cite{CHARM-II:1994aeb,LSND:2001akn,DONUT:2001zvi},
reactors \cite{TEXONO:2009knm}
($1\,{\rm MeV} < E_\nu < 50$\,MeV), and the Sun
(Super-K and Borexino with $0.4\,{\rm MeV} < E_\nu < 20$\,MeV
which is larger than the one
$2\,{\rm keV} < E_\nu < 400$\,keV at DM experiments)
\cite{Super-Kamiokande:2004wqk,Borexino:2017fbd,XENON:2020rca,PandaX-II:2020udv,AtzoriCorona:2022jeb}.
The neutrino decay observations can also put constraints
by searching the decay photons in reactors
\cite{Raffelt:1999gv} or distortions to the CMB
spectrum \cite{Mirizzi:2007jd}, reaching
$10^{-5}\mu_B$ to $10^{-8}\mu_B$. In addition, the stellar
cooling due to plasmon decay
($\gamma^*\rightarrow \nu \bar \nu$) constraints the
magnetic/electric moment (charge radius/anapole) to
$\mathcal O(10^{-12}\,\mu_B)$ ($\mathcal O(10^{-32})$\,cm$^2$)
\cite{MillerBertolami:2014oki,Hansen:2015lqa,Diaz:2019kim}.

However, all existing searches presented above have
intrinsic caveats. With neutrinos not being detected
and especially difficult to disentangle their mass
eigenstates, constraints are obtained as a
combination of parameters involving both the neutrino
mixing and form factor matrix elements \cite{Grimus:1997aa}
to introduce various degeneracies
\cite{AristizabalSierra:2021fuc}. A single or just a handful
of observations cannot disentangle and identify individual
form factor elements. One possible solution
\cite{Ge:2022cib} is the
radiative emission of neutrino pair (RENP) from excited atoms 
\cite{Yoshimura:2006nd,Dinh:2012qb,Fukumi:2012rn,Dinh:2014toa,Yoshimura:2015yja,
Huang:2019phr}.
Since atomic transition energies are typically
$\mathcal O({\rm eV})$, close to the 
neutrino mass values, RENP is an ideal place to test
the light mediators that couples with neutrino and
electrons \cite{Ge:2021lur}, including the massless photon.

In this work we further show that RENP can disentangle
all the neutrino EM properties ($\mu_\nu$, $\epsilon_\nu$,
$\langle r^2_\nu\rangle$, and $a_\nu$) with details.
In \gsec{sec:EMinteractions}, we describe the 
general electromagnetic vertex properties
to establish the formalism of neutrino EM form factors
and summarize the existing experimental/observational
constraints. The selection rules and details of how
neutrino EM properties affect the RENP photon spectrum
are presented in \gsec{sec:renp_EM}. Based on the
theoretical formulas, we illustrate a practical
strategy of scanning the photon spectrum to identify
each form factor elements and their type in
\gsec{sec:RENPsensitivity}.
Our discussions and conclusion can be found in
\gsec{sec:Conclusion}.

\section{Neutrino Eletromagnetic Properties}
\label{sec:EMinteractions}

\subsection{Electromagnetic Form Factors}

The neutrino EM interaction 
is characterized as the neutrino current 
($j_\mu^{(\nu)}$) in terms of a 
vertex function ($\Lambda_\mu^{ij}
(q)$),
\begin{equation}
  j_\mu^{(\nu)}
\equiv
  \bar u(p_i) \Lambda_\mu^{ij}(q)  u(p_j),
\label{eq:nuEMcurrent}
\end{equation} 
with the momentum transfer $q \equiv p_i - p_j$
as difference between the initial ($p_i$) and final
($p_j$) neutrino momenta. Since the neutrino current
vertex $\Lambda_\mu^{ij}(q)$
is a function of the momentum transfer $q$, it
can be expanded as a linear combination of $q_\mu$,
$\gamma_\mu$ and $\sigma_{\mu \nu} q^\nu$ to match
the current Lorentz structure. Each type can be
associated with a $\gamma_5$ to double the form 
factors $F_A^{ij}(q^2)$ with $A= 1, \dots, 6$,
\cite{Giunti:2014ixa},
\begin{align}
  \Lambda_\mu^{ij}
\equiv
  q_\mu (F_1^{ij} + F_2^{ij} \gamma_5)
+ 
  \gamma_\mu  (F_3^{ij}  + F_4^{ij} \gamma_5)
+ 
  i\sigma_{\mu\nu} q^\nu 
  (F^{ij}_5  + F^{ij}_6 i\gamma_5).
\label{eq:Lambda_General}
\end{align}
Then the coupling with a massless photon takes
the form as $\mathcal L \ni A^\mu j^{(\nu)}_\mu$.

Note that the quantum electrodynamics (QED)
gauge symmetry $U(1)_{\rm em}$
is not broken even in the presence of neutrino EM
interactions. Consequently, the $U(1)_{\rm em}$
gauge invariance implies the conservation of 
the neutrino current, $q^\mu  j_{\mu}^{(\nu)} = 0$,
which restricts two of the 6 form factors.
Of the 6 form factors, the first two
$F^{ij}_1 = - F^{ij}_3 (m_i - m_j) / q^2$ and
$F^{ij}_2 = - F^{ij}_4 (m_i + m_j) q^2$ are
expressed in terms of $F^{ij}_3$ and $F^{ij}_4$.
It is interesting to observe that $F^{ij}_1$ is
nonzero only for $m_i \neq m_j$. In other words,
$F^{ij}_1$ has only off-diagonal elements. For
the diagonal case, only $F^{ij}_3$ survives
and the current conservation holds with vanishing
$\bar u_i \slashed q u_j = (m_i - m_j) \bar u_i u_j$
and $i = j$. Note that the current conservation
does not require on-shell photon but on-shell
fermions. Then \geqn{eq:Lambda_General} reduces to,
\begin{eqnarray}
  \Lambda_\mu^{ij}(q)
\equiv
  \left(
    \gamma_\mu - \frac{q_\mu \slashed q}{q^2}
  \right)
  \left[
    f_Q^{ij}(q^2) 
  +
    f_A^{ij}(q^2)
    q^2
    \gamma_5
  \right]
- 
  f_M^{ij}(q^2)i \sigma_{\mu\nu} q^\nu 
+ 
  f_E^{ij}(q^2)\sigma_{\mu \nu} q^\nu \gamma_5
  \,.
  \label{eq:EMvertex}
\end{eqnarray}
To make the physics picture more transparent,
we have redefined, $f^{ij}_Q \equiv F^{ij}_3$ (electric charge),
$f^{ij}_A \equiv F^{ij}_4 / q^2$ (anapole),
$f^{ij}_M \equiv - F^{ij}_5$ (magnetic dipole),
and $f^{ij}_E \equiv F^{ij}_6$ (electric dipole),
respectively. The hermiticity of the interaction Lagrangian 
further requires that $(f_B^{ij})^\dagger = 
f_B^{ij}$.

The form factors at $q^2 \rightarrow 0$
give the values of the magnetic 
moment $\mu_\nu^{ij} \equiv f^{ij}_M(0)$, 
electric dipole $\epsilon_\nu^{ij} 
\equiv f^{ij}_{E}(0)$, charge $q_\nu^{ij} 
\equiv f^{ij}_Q(0)$, and anapole $a_\nu^{ij} 
\equiv f^{ij}_A(0)$. In the zero momentum transfer limit,
these form factors have vanishing effect with the
only exception of the electric charge while the other
three are suppressed by either $q^2$ or $q^\nu$.
Consequently, the neutrino charge is highly constrained,
$|q_\nu| < 3\times 10^{-21}e$ 
\cite{Raffelt:1999gv} and the next order expansion 
of $f^{ij}_Q$ provides the charge radius, 
$\langle r^2_\nu \rangle \equiv  6 
\left. d f^{ij}_Q/dq^2 \right|_{q^2 = 0}$.

For Majorana neutrinos, the current in 
\geqn{eq:nuEMcurrent} contains an extra 
term. While Dirac neutrino fields are 
formed with two sets of operators $a$ and $b$, 
$\nu^{(D)} \sim a u + 
b^\dagger v$, the Majorana case 
has only one $a$, $\nu_k^{(M)} \sim a u + 
a^\dagger v$. A general Majorana interaction 
$\mathcal L = \frac 1 2 \bar \nu_i 
\Lambda_\mu^{ij}A^\mu \nu_j$ generates 
two contributions to the neutrino current,
\begin{equation}
  j_\mu^{(\nu)} 
= 
  \frac 1 2 
\left[ 
  \bar u(p_i) \Lambda_\mu^{ij}(q) u(p_j)
- \bar v(p_j) \Lambda_\mu^{ji}(q) v(p_i)
\right],
\end{equation} 
with switched indices $i$ and $j$.
The charge conjugation transformation matrix $C$
in the spinors, $\overline u = v^T C^{-1}$,
is used to swap $p_i\leftrightarrow p_j$ 
in the above equation to,
\begin{equation}
  j_\mu^{(\nu)} 
= 
 \bar u(p_i) 
 \left[ 
  \frac{
    \Lambda_\mu^{ij}(q)
  +
    C (\Lambda_\mu^{ij}(q))^T C^{-1}
  }{2}
 \right] 
  u(p_j)
  \,.
\end{equation} 
Since the charge conjugate transformation 
of the Lorentz structures follows, 
\begin{eqnarray}
  C \Gamma^T C^{-1}
=
-
  \Gamma 
\,\quad {\rm for} \quad 
\Gamma = \gamma_\mu, \sigma_{\mu \nu}, 
\sigma_{\mu \nu}\gamma_5
\,,
\quad {\rm and} \quad 
  C (\gamma_\mu \gamma_5)^T C^{-1}
=
  \gamma_\mu \gamma_5
  \,,
\end{eqnarray}
the extra contribution for Majorana neutrinos 
does not mix Lorentz structures and introduces 
at most a minus sign.

The only difference 
between Dirac/Majorana neutrinos with EM 
interactions is the form factor matrix pattern,
\begin{eqnarray}
  (q_{\nu}^{ij})^T 
=
  - q_{\nu}^{ij}
\,, \quad
  (\mu_{\nu}^{ij})^T 
=
  - \mu_{\nu}^{ij}
\,, \quad
  (\epsilon_{\nu}^{ij})^T 
=
  - \epsilon_{\nu}^{ij}
\,, \quad {\rm and} \quad 
  (a_\nu^{ij})^T
=
  a_\nu^{ij}
  \,.
\end{eqnarray}
In other words, some form factors are forbidden for
Majorana neutrinos and consequently the number of independent
parameters is significantly reduced. Observing nonzero
diagonal element $q^{ii}_\nu$, $\mu^{ii}_\nu$, or
$\epsilon^{ii}_\nu$ is a direct evidence of the Dirac
nature of neutrinos. In addition, the existence of
electric dipole moment $\epsilon^{ij}_\nu$
indicates possible CP violation. For comparison,
anapole violates both parity and charge conjugations,
but preserves CP \cite{Giunti:2014ixa}.

\subsection{Existing Terrestrial and Celestial Tests}
\label{sec:current_bounds}

The constraints on the neutrino EM form factors 
are obtained from a variety of terrestrial experiments and 
astrophysical observations \cite{Giunti:2014ixa,Giunti:2022aea}.
Nevertheless, these existing measurements can only
constrain combinations of the relevant form factor
elements. In other words, there is no way to distinguish
and identify the individual ones.

\subsubsection{Terrestrial Tests with Electron Scattering}

The terrestrial experiments typically use neutrino
scattering with electron or nuclei via a $t$-channel
photon exchange to probe the neutrino EM form factors.
Either reactor, accelerator, or solar neutrinos are
possible sources for such experiments.
Currently, the strongest constraint comes from the
neutrino scattering with electron. Such process can already
be induced by the SM weak interactions with differential
cross section \cite{Kayser:1979mj,Rodejohann:2017vup},
\begin{eqnarray}
  \left( 
  \frac{d\sigma_{\nu_i e\rightarrow \nu_j e}}{dT_r} 
  \right)_{\rm SM}
\hspace{-2mm}
= 
\frac{G_F^2 m_e}{2 \pi}
\Biggl\{
  |v_{ji} \mp a_{ji}|^2
+
  |v_{ji} \pm a_{ji}|^2
  \left(1 - \frac{T_r}{E_{\nu_i}} \right)^2
\hspace{-2mm}
-
\left( 
  \left|
   v_{ji} 
  \right|^2
-
  |a_{ji}|^2
\right) 
 \frac{m_e T_r}{4E_{\nu_i}^2}
\Biggr\},
\label{eq:xsec_nueTOnue}
\end{eqnarray}
where the upper (lower) sign is for the
(anti-)neutrino scattering. The SM neutrino
and electron EW interactions have been summarized
into the axial and vector current couplings
$a_{ij} \equiv  - U_{ei}^* U_{ej} + \delta_{ij}/2$ 
and $v_{ij} \equiv - a_{ij} + 2 \sin \theta_w\delta_{ij}$
of the effective four-fermion operators,
$\bar e \gamma_\mu (v_{ij} + a_{ij} \gamma_5) e
\times \bar \nu_i \gamma^\mu (1 - \gamma_5) \nu_j$.
Both the charged and neutral currents mediated by
the $W^\pm$ and $Z$ gauge bosons can contribute.
The neutrino mixing matrix elements $U^*_{ei}$ and
$U_{ej}$ arise from the charged current contribution
with the flavor index $e$ for the involved electron
in the initial and final states. Also,
the neutrino external states are in their physical mass
eigenstates. In other words, \geqn{eq:xsec_nueTOnue}
considers the incoming neutrinos as incoherent states.
The more general case with coherence, especially for
neutrino oscillation experiments, is elaborated below.

Non-zero neutrino charge radius \cite{Grau:1985cn}
and anapole \cite{Khan:2022akj} modify the effective
vector current coupling $v_{ji} \rightarrow \widetilde v_{ji}$
as,
\begin{eqnarray}
  \widetilde v_{ij}  
\equiv 
  v_{ij} 
+ 4
\left[ 
  \frac{\langle r^2_\nu \rangle_{ij}}{6} 
- (a_\nu)_{ij}
\right]
  \frac{M_W^2 \sin^2 \theta_w}{e}.
\end{eqnarray}
Although the charge randius and anapole are associated
with the neutrino vector and axial currents, respectively,
their contributions can altogether combine into $v_{ij}$
but not $a_{ij}$. Note that $v_{ij}$ and $a_{ij}$ are
actually associated with the electron bilinears in the
effective four-fermion operators, not the neutrino ones.

Since both charge radius and anapole appear in the
same place of $\widetilde v_{ij}$, an unfortunate
degeneracy between them exists. Making it worse,  
the incoming neutrinos are typically coherent linear
combination of mass eigenstates and oscillate over distance
while the final-states neutrinos should be treated
as the physical mass eigenstates. Being illusive, all
the final-state
neutrinos $\nu_j$ contribute to the electron scattering
event rate as an inclusive cross section. 
The actual observable is a summation over the indices 
$i$ and $j$ in \geqn{eq:xsec_nueTOnue} with coefficients given 
by the neutrino mixing matrix elements \cite{Grimus:1997aa}. 
In addition, the tiny neutrino masses are negligible and
different final-state neutrinos have no actual difference.
These lead to an accidental coincidence that the final-state
neutrinos can also be taken as flavor states.
For short baseline, no oscillation occurs and the couplings 
are written in the flavor basis, $\widetilde v_{\alpha \beta} 
\equiv \sum_{ij} U_{\alpha i} U_{\beta j}^* \widetilde v_{ij}$. 
Since $a_{\alpha \beta} = - \delta_{e \alpha} \delta_{e \beta} + \delta_{\alpha \beta} / 2$
and $v_{\alpha \beta} = - a_{\alpha \beta} + 2 \sin \theta_w
\delta_{\alpha \beta}$ are diagonal, the differential
cross section in \geqn{eq:xsec_nueTOnue} becomes,
\begin{align}
\hspace{-3mm}
    \frac{d\sigma_{\nu_\alpha e\rightarrow X e}}{d T_r} 
& =
\frac{G_F^2 m_e}{2 \pi}
\Biggl\{
  | \widetilde v_{\alpha \alpha} \mp a_{\alpha \alpha}|^2
+
  | \widetilde v_{\alpha \alpha} \pm a_{\alpha \alpha}|^2
  \left(1 - \frac{T_r}{E_\nu} \right)^2
-
\left( 
  |\widetilde v_{\alpha \alpha}|^2
- |a_{\alpha \alpha}|^2
\right) 
 \frac{m_e T_r}{4 E_\nu^2}
\Biggr\}
\nonumber
\\
& + 
  \frac{\alpha m_e}{144}
  \sum_{\beta\neq \alpha} 
    |\langle r^2_\nu \rangle_{\alpha \beta}^{\rm eff} |^2
  \left[
    1 
  +
    \left(1 - \frac{T_r}{E_{\nu}} \right)^2
  -
   \frac{m_e T_r}{4E_{\nu}^2}
  \right]
  \label{eq:Xsec_nuaeToXe}
  \,.
\end{align}
Note that the diagonal elements
$\langle r^2_\nu \rangle_{\alpha \alpha}$
and $(a_\nu)_{\alpha \alpha}$ appear in the first
line while the off-diagonal ones in the second line.
Although the diagonal elements have interference
with the SM contribution, the off-diagonal ones
are standalone as square term in the second line.

For convenience, we have defined the effective charge
radius as,
\begin{eqnarray}
  \langle r^2 \rangle_{\alpha\beta}^{\rm eff}
\equiv 
  \sum_{ij} 
  U_{\alpha i}
  U_{\beta j}^* 
  \left[ 
    \langle r_\nu^2 \rangle_{ij} 
  -
    6 (a_\nu)_{ij}
  \right]
  \,.
\end{eqnarray}
With degeneracy among the charge radius, anapole, and the
neutrino mixing matrix elements $U_{\alpha i} U^*_{\beta j}$, 
it is very hard to disentangle and identify the 
individual neutrino EM form factors.

A similar situation occurs for the magnetic ($\mu_\nu$) and 
electric dipole ($\epsilon_\nu$) moments. The cross-section in 
\geqn{eq:xsec_nueTOnue} receives corrections as
\cite{Vogel:1989iv},
\begin{eqnarray}
  \left( 
  \frac{d\sigma_{\nu_\alpha e\rightarrow X e}}{dT_r} 
  \right)_{\mu,\epsilon}
=
  \frac{\pi\alpha^2}{m_e^2}
  \left(
    \frac1 {T_r}
  -
    \frac1{E_{\nu_\alpha}}
  \right)
  |\mu^{\rm eff}_{\alpha}|^2,
\label{eq:dXsec-ME}
\end{eqnarray}
with $X$ denoting any possible neutrino flavor or mass
eigenstates. The effective magnetic moment
$|\mu^{\rm eff}_{\alpha}|^2$ is
defined as combination of the magnetic ($\mu_\nu$)
and electric dipole ($\epsilon_\nu$) moments,
\begin{align}
  |\mu^{\rm eff}_{\alpha}|^2
& \equiv
  \sum_{ij} 
\left\{
    U_{\alpha i}
    (\mu^2_\nu + \epsilon^2_\nu)_{ij}
    U_{\alpha j}^* 
  +
    2 \, {\rm Im}
    [
      U_{\alpha i}
      (\mu_\nu \epsilon_\nu)_{ij}
      U_{\alpha j}^* 
    ]    
  \right\}.
\end{align}
When defining the effective MDM $\mu^{\rm eff}_\alpha$,
we have used 
the property that $\mu^\dagger_\nu = \mu_\nu$ and
$\epsilon^\dagger_\nu = \epsilon_\nu$ are hermitian.
Since $(\mu_\nu)_{ij}$ and $(\epsilon_\nu)_{ij}$
appear in similar ways, the neutrino scattering
process cannot distinguish them.
Multiple degeneracies exist
not just for the charge radius and anapole, but 
also EDM and MDM. Although there are four
form factors each with 6 elements, only few 
independent combinations can be measured.

The bound on the effective charge radius is
obtained for only the diagonal elements,
assuming vanishing non-diagonal ones
\cite{Kouzakov:2017hbc}. The most stringent bounds 
at 90\% C.L. come from TEXONO \cite{TEXONO:2009knm} 
with $\bar \nu_e$ scattering and CHARM-II with 
$\nu_\mu$ scattering \cite{CHARM-II:1994aeb},
\begin{eqnarray}
  -4.2\times 10^{-32} {\rm cm}^2
<
  \langle r^2_\nu \rangle_{ee}^{\rm eff}
< 
  6.6 \times 10^{-32} {\rm cm}^2,
\quad {\rm and} \quad 
  |\langle r^2_\nu \rangle_{\mu\mu}^{\rm eff}|
< 
  1.2 \times 10^{-32}
  {\rm cm}^2,
  \label{eq:R2eff_bound_escattering}
\end{eqnarray}
respectively. On the other hand, the strongest
constraints on the effective MDM
$\mu^{\rm eff}_{\alpha}$,
comes from the GEMMA $\bar \nu_e$ 
scattering \cite{Beda:2012zz}, LSND $\nu_\mu$ 
scattering \cite{LSND:2001akn}, and DONUT 
$\nu_\tau$ scattering \cite{DONUT:2001zvi},
\begin{eqnarray}
  |\mu^{\rm eff}_{e}|
<
  2.9\times 10^{-11}\mu_B,
\quad
  |\mu^{\rm eff}_{\mu}|
<
  6.8\times 10^{-10}\mu_B,
\quad {\rm and} \quad
  |\mu^{\rm eff}_{\tau}|
<
  3.9\times 10^{-7}\mu_B,
  \label{eq:mueff_bound_escattering}
\end{eqnarray}
at 90\% C.L. All these 5 experiments
have short baselines with no need of considering the
neutrino oscillation effect. Generally speaking, the
neutrino oscillation effect should be taken into
consideration altogether \cite{Kouzakov:2017hbc}.

\subsubsection{Terrestrial Tests with Nuclei Scattering}

With low momentum transfer ($E_\nu < \mathcal O(10)$\,keV),
neutrinos interact coherently with atomic nucleus
and the cross section is enhanced by the number of
nucleons squared. In addition, the differential
cross section receives one more enhancement from
the photon propagator $1/q^2$ for small $q^2$.
With larger cross-section, the
coherent elastic neutrino-nucleons scattering 
(CE$\nu$NS) is a promising probe for neutrino EM properties.
The CE$\nu$NS cross-section with EM interactions is
slightly different from the electron one in
\geqn{eq:Xsec_nuaeToXe} and \geqn{eq:dXsec-ME}
with just neutral current
in the SM \cite{Cadeddu:2018dux},
\begin{align}
   \left(\frac{d\sigma}{dT_r}\right)_{\rm CE\nu NS}
& =
  \frac{G_F^2M}{2\pi}
  \left( 
    1
  -
    \frac{MT_r}{2 E_\nu^2} 
  \right) 
  \left[ 
    \widetilde v_{\alpha \alpha}^{~p} 
    Z F_p
  +
    \widetilde v_{\alpha \alpha}^{~n}
    N F_n  
  \right]^2
\\ & \nonumber 
+
    \frac{\alpha Z^2 F_p^2}{144}
      \sum_{\beta\neq \alpha} 
    |\langle r^2_\nu \rangle_{\alpha \beta}^{\rm eff} |^2
    \left( 
      1
    -
      \frac{MT_r}{2 E_\nu^2} 
  \right) 
+
  Z^2 F_p^2
  \frac{\pi\alpha^2}{m_e^2}
  \left(
    \frac1 {T_r}
   -
    \frac1{E_{\nu_\alpha}}
  +
    \frac{T_r}{4 E_{\nu_\alpha}^2}
  \right)
  |\mu^{\rm eff}_{\alpha}|^2.
\qquad
\end{align}
where $Z$ $(N)$ is the number of protons (neutrons) and 
$F_{p}$ ($F_n$) the proton (neutron) form factor. 
Similar to \geqn{eq:Xsec_nuaeToXe}, only the diagonal 
charge radius elements have interference with the
proton/neutron vector ($v_{\alpha \alpha}^{p} = 1/2 - 2 \sin^2 
\theta_w$ and $v_{\alpha \alpha}^{n} = -1/2$) couplings through 
$\widetilde v_{\alpha \alpha}^{~p,n} = v_{\alpha \alpha}^{p,n} + 2 
\langle r_\nu^2 \rangle_{\alpha \alpha}M_W^2 \sin^2 \theta_w/3e$.
The axial current contribution is proportional to the 
difference between nucleons with spin-up and spin-down
which is typically small for the atoms in CE$\nu$NS 
experiments and hence can be neglected \cite{Barranco:2005yy}.
Clearly, the CE$\nu$NS constraints suffer from the
parameter degeneracy discussed before.
The COHERENT experiment bounds \cite{COHERENT:2020iec} 
on the diagonal matrix elements $\langle r^2_\nu \rangle_{\alpha 
\alpha}^{\rm eff}$ and $\mu^{\rm eff}_{\alpha}$ are 
typically 1$\sim$2 orders of magnitude larger than those in 
\geqn{eq:R2eff_bound_escattering} 
and \geqn{eq:mueff_bound_escattering} \cite{Cadeddu:2018dux,Khan:2022akj},
\begin{eqnarray}
  |\langle r^2_\nu \rangle_{ee}^{\rm eff}|,\,
  |\langle r^2_\nu \rangle_{\mu\mu}^{\rm eff}|
<
  6 \times 10^{-31} {\rm cm}^2
\,,\quad 
  |\mu^{\rm eff}_{e}|
< 
  4\times 10^{-9}\mu_B,
\quad {\rm and} \quad 
  |\mu^{\rm eff}_{\mu}|
<
  4\times 10^{-10}\mu_B
  ,
\end{eqnarray}
at 90\% C.L. and for the transition matrix elements,
\begin{eqnarray}
  |\langle r^2_\nu \rangle_{e\mu}^{\rm eff}|
< 
  3.3 \times 10^{-31}
  {\rm cm}^2
\quad {\rm and} \quad 
  |\langle r^2_\nu \rangle_{e\tau}^{\rm eff}|,  
  |\langle r^2_\nu \rangle_{\mu\tau}^{\rm eff}|
< 
  4 \times 10^{-31}
  {\rm cm}^2.
\end{eqnarray}

\subsubsection{Solar Neutrino Measurements with Electron Scattering}

In addition to man-made sources, the solar neutrinos
can also be used to measure the neutrino EM properties.
With nontrivial transition probability
$P^{~\rm sun}_{\nu_e \rightarrow \nu_i}(E_\nu) 
= |\widetilde U_{e i}|^2$ from the electron neutrino
in the solar interior to incoherent mass eigenstates
when arriving the solar surface, as arising from
the MSW effect \cite{Wolfenstein:1977ue,
Mikheyev:1985zog,Smirnov:2019kto} with adiabatic evolution \cite{Parke:1986jy},
the effective MDM $\mu^\odot_{\nu}$ and the effective 
charge radius $\langle r_\nu^2\rangle^\odot$
become dependent on the
neutrino energy through the effective mixing
matrix $\widetilde U_{e i}$ at the production places
inside the Sun \cite{Grimus:2002vb},
\begin{eqnarray}
  (\mu^\odot_{\nu})^2
\equiv
  \sum_{ij}
  |\widetilde U_{ej}|^2
  |(\mu_{\nu})_{ij} - i (\epsilon_{\nu})_{ij}|^2
\quad {\rm and} \quad 
  \langle r_\nu^2\rangle^{\odot} 
\equiv 
  \sum_{ij} 
  |\widetilde U_{ej}|^2
  |
    \langle r_\nu^2 \rangle_{ij}
  -
    6(a_\nu)_{ij}
  |
  .
\end{eqnarray}
With non-uniform matter density inside the Sun,
the effective mixing matrix $\widetilde U_{ei}$
and hence the transition probability are radius
dependent functions. Consequently, the effective
mixing matrix should contain an integration
over the solar radius weighted by the matter density.
However, the matter effect either ceases for
low-energy neutrinos ($E_\nu \lesssim 1$\,MeV), with 
$\widetilde \theta_{12} = \theta_{12}$, or dominates
for high-energy neutrinos ($E_\nu \gtrsim 5$\,MeV), 
with $\widetilde \theta_{12} = \pi/2$.
It is then more convenient to use different values 
for the effective magnetic moment in individual 
regions. The Borexino experiment \cite{Borexino:2017rsf} 
is sensitive to the $pp$ and $^{7}$Be neutrino spectra 
at $E_\nu \lesssim 0.42$ and $0.86$\,MeV, respectively,
to give constraints
\cite{Khan:2017djo,Borexino:2017fbd},
\begin{eqnarray}
  |\mu^\odot_{\nu}| 
<
  2.8\times 10^{-11}\mu_B
\quad {\rm and} \quad 
  - 0.82 \times 10^{-32}\,{\rm cm}^2 
<
  \langle r_\nu^2\rangle^{\odot} 
<
   1.27 
   \times 10^{-32}\,{\rm cm}^2,
\end{eqnarray}
at 90\% C.L. The Super-Kamiokande (SK) 
experiment has a 5\,MeV energy threshold
and measures the $^8B$ neutrino flux
with $E_\nu \in [5,20]$\,MeV 
\cite{Super-Kamiokande:2004wqk} to obtain,
\begin{eqnarray}
  |\mu^\odot_{\nu}| 
<
  1.1\times 10^{-10}\mu_B
\quad {\rm and} \quad 
  |\langle r_\nu^2\rangle^{\odot}|
<
  6.9\times10^{-32}\,{\rm cm}^2,
\end{eqnarray}
at $90\%$\,C.L \cite{Super-Kamiokande:2004wqk,Joshipura:2001ee}.

Since the solar neutrino flux is much lower than
the man-made one, larger detector (than the kilogram
detector used in the CE$\nu$NS experiment) is needed.
In addition, with the $pp$ flux dominating at low
energy ($E_\nu < 400$\,keV), a detector with lower 
energy threshold is better.
These two factors renders an opportunity for
the dark matter direct detection experiments
to probe neutrino EM properties.
While the Xenon1T experiment claimed a likely signal
of the magnetic moment $\mu_\nu^{\odot} = 2.2\times 10^{-11}\mu_B$
at $2.7\,\sigma$ level \cite{XENON:2020rca}
in the June of 2020,
PandaX-II could not confirm but put a stringent 
bound $<4.9\times 10^{-11}\mu_B$ at 90\% C.L.
\cite{PandaX-II:2020udv}. The bound is further improved
by the recent XENONnT \cite{XENON:2022ltv} and
LUX-Zepelin \cite{AtzoriCorona:2022jeb} measurements,
\begin{eqnarray}
  |\mu^{~\rm \odot}_{\nu}|
<
  6.4\times 10^{-12}\mu_B,
\quad {\rm and} \quad 
  |\mu^{~\rm \odot}_{\nu}|
<
  1.1\times 10^{-11}\mu_B 
\quad {\rm at} \quad  90\%\,{\rm C. L.},
\end{eqnarray}
respectively. While the strongest constraints for the charge
radius comes from PandaX-II and XENONnT\cite{Khan:2022bel},
\begin{eqnarray}
  |\langle r_\nu^2\rangle^{\odot}|
<
  6\times 10^{-29}\,{\rm cm}^2,
\quad {\rm and} \quad 
  -4.5\times 10^{-31}\,{\rm cm}^2
<
  \langle r_\nu^2\rangle^{\odot}
<
  3\times 10^{-32}\,{\rm cm}^2,
\end{eqnarray}
at $90\%$\,C.L.

\subsubsection{Neutrino Electromagnetic Decay}

With non-zero neutrino masses, the presence of EM interactions 
allows for neutrino visible decay into photon. For $m_i > m_j$,
the reaction $\nu_i \rightarrow \nu_j + \gamma$ may occur. 
Since photons are on their mass-shell $q^2 = 0$,
only $\mu_\nu$ and $\epsilon_\nu$ that are proportional to
a single $q$ can contribute while the charge radius and
anapole with $q^2$ suppression has vanishing effect.
The decay rate \cite{Raffelt:1999gv},
\begin{eqnarray}
    \Gamma(\nu_i\rightarrow \nu_j + \gamma)
=
  \frac{1}{8\pi} 
  \left(\frac{m_i^2 - m_j^2}{m_i}\right)^3
  \left[
    |(\mu_\nu)_{ij}|^2
  +|(\epsilon_\nu)_{ij}|^2
  \right],
\end{eqnarray}
is a sum of matrix elements squared. So the constraint
on individual element is not relaxed from the combined one.
However, the constraints from observing neutrino decay is
not that strong.

The reactor experiments put upper bounds on the decay width 
from the absence of decay photons while the
solar neutrino flux reduces in the presence 
of neutrino decay \cite{Raffelt:1999gv}.
For the effective magnetic moment $|\mu^{\rm dec}_\nu|^2 
\equiv |(\mu_\nu)_{ij}|^2 + |(\epsilon_\nu)_{ij}|^2$
that appears in the decay process,
the bounds in \cite{Raffelt:1999gv} are,
\begin{eqnarray}
   |\mu^{\rm dec}_\nu|
<
  0.9\times 10^{-1}
  \left(\frac{{\rm eV}}{m_\nu} \right)^2
  \mu_B,
\quad {\rm and} \quad 
  |\mu^{\rm dec}_\nu|
<
  0.5\times 10^{-5}
  \left(\frac{{\rm eV}}{m_\nu} \right)^2
  \mu_B,
\end{eqnarray}
at 90\% C.L. for reactor and solar neutrinos respectively.

The decay photons can also distort the CMB
blackbody spectrum which fits the COBE/FIBRAS data quite
nicely with precision at the level of $10^{-4}$
\cite{Fixsen:1996nj} or equivalently
\cite{Mirizzi:2007jd},
\begin{align}
  |\mu^{\rm dec}_\nu| 
< 3\times 10^{-8}\mu_B ~({\rm NO}) 
\,,\quad {\rm and} \quad 
  |\mu^{\rm dec}_\nu| 
< 3.5\times 10^{-7}\mu_B ~({\rm IO}),
\end{align}
for the normal (NO) and inverted (IO) mass orderings,
respectively.

\subsubsection{Celestial Tests with Stellar Cooling}

The strongest constraints for neutrino EM properties 
are obtained from astrophysical observations. The stellar
cooling caused by plasmon decay ($\gamma^*\rightarrow \bar \nu 
\nu)$ are sensitive to the combination 
\cite{Haft:1993jt},
\begin{eqnarray}
  (\mu^{\ostar}_{\nu})^2
\equiv
  \sum_{ij}
\left\{
   2m_{\gamma^*}^2 
\left[
   \frac{|\langle r^2_\nu\rangle_{ij}|^2}{36}
  +|(a_\nu)_{ij}|^2
\right]
+ |(\mu_{\nu})_{ij}|^2 
+ |(\epsilon_{\nu})_{ij}|^2
  \right\}
  ,
\end{eqnarray}
where $m_{\gamma^*} \sim 10$\,keV is the plasmon 
mass \cite{Haft:1993jt}.
The constraint from the red giant cooling is
$|\mu^{\ostar}_{\nu}| < 2.2\times 10^{-12}\mu_B$ 
at 90\% C.L. \cite{Diaz:2019kim} while the white dwarf
cooling provides $|\mu^{\ostar}_{\nu}| < 2.9\times 
10^{-12}\mu_B$ at 90\% C.L. 
\cite{MillerBertolami:2014oki,Hansen:2015lqa}. These
numbers convert to
$\sqrt{|\langle r^2_\nu\rangle_{ij}|^2 / 36
  +|(a_\nu)_{ij}|^2} < (1.8 \sim 2.4) \times 10^{-32}$\,cm$^2$ for the charge radius and
  anapole.
Although the constraint on individual elements
cannot be relaxed with sum of squared terms,
these numbers are subject to large theoretical
uncertainty for the complicated environments inside
a compact star \cite{Stancliffe:2016sa}.

\subsubsection{Cosmological Constraints}

Cosmological observations can also constrain the electromagnetic
moments for Dirac neutrinos. The right-handed 
neutrinos can be thermally produced in the presence 
of nonvanishing $\mu_\nu$ or $\epsilon_\nu$
in the early Universe and modify the effective
relativistic degrees of freedom, $N_{\rm eff}$.
The bounds range from $|(\mu_{\nu})_{ij}|$, 
$|(\epsilon_{\nu})_{ij}| < 2.7\times 10^{-12}\mu_B$ 
to  $|(\mu_{\nu})_{ij}|$, $|(\epsilon_{\nu})_{ij}|
< 3\times 10^{-11}\mu_B$ if 3 to 1 neutrinos 
are produced, respectively 
\cite{Li:2022dkc}. Since no right-handed neutrinos
can be produced without changing the neutrino
chirality, this constraint does not
apply for the charge radius or anapole.

\subsubsection{Caveats}

All bounds available have caveats. The cosmological 
bounds are valid for Dirac neutrinos only while
the stellar cooling bounds suffer from uncertainties
in the star modeling
\cite{Stancliffe:2016sa} and can be evaded in 
the presence of BSM light particles \cite{Babu:2020ivd}.
Finally, all existing constraints are sensitive 
to combinations of EM parameters and even the
neutrino mixing
matrix elements. In other words, they suffer from
parameter degeneracies
\cite{AristizabalSierra:2021fuc}.
As we elaborate below, the RENP process of atomic
transitions can provide a unique way of probing
the individual elements of the neutrino EM form
factors. Once experimentally realized, RENP would
become a powerful tool for measuring the neutrino
properties.

\section{RENP with Neutrino EM Form Factors}
\label{sec:renp_EM}

As elaborated in \gapp{App:Htot}, the total
Hamiltonian characterizing the RENP process has
three parts,
\begin{eqnarray}
  H 
=
  H_0 + H_\gamma + H_{\nu \bar\nu},
\label{eq:Total_H}
\end{eqnarray}
where $H_0$ describes the atomic energy levels such that 
$H_0|a\rangle = E_a |a\rangle$ with $a = v,e,g$. The RENP
process arises in the SM from the single-photon emission
Hamiltonian $H_\gamma$ and the neutrino pair emission
Hamiltonian $H_{\nu \bar \nu}$ through the weak interactions
mediated by the $W/Z$ bosons. An electron is first pumped from
an excited state $| e \rangle$ to an intermediate
virtual state $| v \rangle$ via $H_{\nu\bar \nu}$ by emitting
a pair of neutrinos ($| e \rangle \rightarrow | v \rangle
+ \nu \bar \nu$) and then jumps to the ground state
$| g \rangle$ via $H_\gamma$ by emitting a photon
($| v \rangle \rightarrow | g \rangle + \gamma$).
The neutrino EM properties allow extra contribution 
to the first transition between the excited and 
virtual states.

The whole process $|e\rangle \rightarrow |g\rangle + \gamma 
+ \nu\bar\nu$ is stimulated using two back-to-back 
lasers at frequencies $\omega$ and 
$\omega'$, with $\omega < \omega'$. While the final-state 
photon is emitted in the same direction as the laser 
with smaller frequency, $E_\gamma = 
\omega$, the neutrino pair is emitted in the 
opposite direction with combined energy,
$E_\nu + E_{\bar \nu} = \omega'$. 
In addition to $E_\gamma = \omega$, stimulation
requires the frequencies to be tuned to match the energy
difference between states, $\omega + \omega' = E_e - E_g$
according to the global energy conservation
\cite{Song:2015xaa,Zhang:2016lqp}.

\subsection{Selection Rules for the QED and Weak Transitions}

The stimulation is only possible if the excited 
state is metastable with a lifetime longer than 
1\,ms \cite{Fukumi:2012rn}. In comparison, a typical single-photon
emission process $|e\rangle \rightarrow | g \rangle + \gamma$
has $\mathcal O(1)$\,ns lifetime \cite{Fukumi:2012rn}.
So the single-photon emission needs to be forbidden 
at the first order (E1 or M1). 
The emission of a single photon changes the total
angular momentum from the initial ($J_e$) to the final
one ($J_g$) by at most one single unity, $J_g - J_e = 0$, $\pm 1$,
with the $J_g = J_e = 0$ transition forbidden.
At the same time, the magnetic quantum number
should also change at most by one unity,
$M_{J_e} - M_{J_g} = 0$, $\pm 1$. The relative
parity between the initial ($\pi_e$) and final
($\pi_g$) states determines if the transition is
E1 type with parity flip ($\pi_g = - \pi_e$) or
M1 type with parity conserved ($\pi_g = \pi_e$).
This is summarized in \gtab{tab:selection_rules}.
For the purpose of forbidding the first-order
E1 and M1 transitions, one may choose atomic
states with either (1) $J_g$ and $J_e=0$ or
(2) $|J_g-J_e|>1$. Two possible atomic candidates
found in the literature are Yb with $J_g = J_e = 0$
and Xe with $J_g = 0$ and $J_e = 2$ \cite{Song:2015xaa}.
\begin{table}[t]
\centering
\begin{tabular}{ccc}
  &           E1           &           M1           \\
\hline
$\Delta J$        &        $0,\pm1$        &        $0,\pm1$        \\
$\Delta M_J$      &        $0,\pm1$        &        $0,\pm1$        \\
Parity            &    $\pi_i = -\pi_f$    &    $\pi_i = \pi_f$     \\
\end{tabular}
\caption{\label{tab:selection_rules}
The selection rules for the E1 and M1 transitions
between the initial $i$ and final $f$ states. In addition,
no transition with $J_{i} = J_{f} = 0$ can occur.
}
\end{table}

As summarized in \gapp{sec:Hemission}, the photon
emission Hamiltonian up to leading terms is,
\begin{align}
  H_\gamma
& = 
- i e \boldsymbol A^{(\gamma)} \cdot [\hat{{\boldsymbol x}}, H_0]
+ \frac{e}{2m_e} \boldsymbol \sigma \cdot 
  (\boldsymbol\nabla \times \boldsymbol A^{(\gamma)}).
\label{eq:Hgamma}
\end{align}
The first term contains the spatial position operator $\hat{\boldsymbol x}$. Being parity odd, the position
operator controls the E1 transitions.
The second term connects the photon magnetic field
($\boldsymbol B = \boldsymbol\nabla \times \boldsymbol A^{(\gamma)}$)
with the spin operator $\hat{\boldsymbol s} \equiv \boldsymbol \sigma / 2$.
Since $\hat {\boldsymbol s}$ is 
parity even, the second term controls the M1 transitions.

The E1 and M1 transitions in QED have quite different strength.
The E1 type photon emission is regulated by the dipole transition 
operator ${\boldsymbol d}_{gv}\equiv e \frac { E_{v} - E_g} \omega \langle g| \hat {\boldsymbol x }| v \rangle$ which is of the order of the Bohr's radius $a_0 = 
(m_e \alpha)^{-1}$ times the energy difference $E_v - E_g$.
For photon emission, the E1 type transition
is typically two orders ($\sim 1/\alpha$ where $\alpha$
is the fine structure constant)
larger than its M1 counterpart
as detailed in \gapp{sec:Hemission}. So, the photon emission
$|v \rangle \rightarrow | g \rangle +\gamma$ is selected
to be of the E1 type to increase statistics. 

Since each emitted neutrino has spin 1/2, the neutrino pair emission 
($|e \rangle \rightarrow |v\rangle + \nu\bar \nu$) also changes 
the total angular momentum by at most 1 unity. Consequently, the 
emission follows the same selection rules as described in \gtab{tab:selection_rules}. With partiy violating weak
interactions, the electron coupling to the neutrino current
contains both vector and an axial-vector components,
$\mathcal L_{e\nu} = \sqrt 2 G_{\rm F} \bar e \gamma^\mu (v + a \gamma_5) e j^{(\nu)}_\mu$.
As derived in \gapp{sec:Hew}, the Hamiltonian responsible for
neutrino pair emission is,
\begin{align}
 H_{\nu\bar\nu} 
& = 
- i \sqrt{2} G_{\rm F}
  \left( 
     v{\boldsymbol j}^{(\nu)}   
   + a j_0^{(\nu)} \boldsymbol \sigma 
  \right)
  \cdot  [\hat{\boldsymbol x}, H_0] 
+\sqrt{2} a G_{\rm F} {\boldsymbol j}^{(\nu)}\cdot \boldsymbol{\sigma}.
\label{eq:Hnunubmain}
\end{align}
The two terms inside the parenthesis are associated
with the dipole operator being proportional to the Bohr radius.
Since the dipole operator has odd parity,
$v {\boldsymbol j}^{(\nu)}$ contributes an E1 type 
transition while $aj^{(\nu)}\boldsymbol \sigma$ an E1$\times$M1 type one.
Without $1/m_e$ suppression, the third term
provides the leading contribution which is M1 type.
For the neutrino pair emission, the E1 transition is
$\sim 3$ orders ($E / m_e \alpha$ with the atomic
energy $E \sim \mathcal O($eV)) of magnitude smaller
than the M1 counterpart.

With E1 dipole transition for the photon emission
and M1 type for the neutrino pair emission being the
optimal choices for increasing statistics, the whole
RENP process should be E1$\times$M1 type. In practice,
the transition types is selected by choosing
the target atom states and the laser frequency tuning. 
This is true for not just the SM RENP process with
QED and EW interactions but also for new physics searches
such as the neutrino EM properties.

\subsection{Transition Matrix Elements}

Let us start from the simplest electromagnetic 
interactions on the electron side which is 
conceptually clearer to set the stage. The electric 
dipole moment operator controls the E1 transition with
a single-photon emission, $|v\rangle \rightarrow |g\rangle + 
\gamma$, between the virtual ($|v \rangle$)
and ground ($| g \rangle$) states. In the Coulomb 
gauge, the electron couples with 
the electric field ($\boldsymbol E$) of the emitted photon
through the electric dipole operator ($\boldsymbol d_{gv}$),
\begin{eqnarray}
   \langle g | H_\gamma |v \rangle
\equiv
  \mathcal M_D e^{- i \omega (t + \hat {\boldsymbol p}_\gamma \cdot {\boldsymbol x})},
\quad \mbox{with} \quad
  \mathcal M_D
\equiv 
  - \boldsymbol d_{gv} \cdot \boldsymbol E
  \label{eq:Me_dipole}
  \,,
\end{eqnarray}
where $k \equiv (\omega, \omega \hat{\boldsymbol p}_\gamma)$ 
is the photon momentum. 

For the SM interactions and Dirac neutrinos,
the neutrino pair emission matrix element is
\cite{Yoshimura:2006nd,Dinh:2012qb,Song:2015xaa,Ge:2022cib},
\begin{eqnarray}
  \langle v | H_{\nu \bar \nu} |e\rangle
\equiv
  \mathcal M_W^{(D)} e^{-i(p_\nu+p_{\bar \nu})\cdot x},
\quad {\rm with} \quad 
  \mathcal M_W^{(D)}
\equiv 
  \sqrt{2}G_{\rm F} 
  a_{ij} 
  (0, {\boldsymbol s}_{ve})^\mu 
~ \bar u(p_{\nu_i})\gamma_\mu P_L v(p_{\bar\nu_j}),
\label{eq:Mw-Dirac}
\end{eqnarray} 
where $\boldsymbol s_{ve} \equiv \langle v | \boldsymbol \sigma | e \rangle / 2$
is the spin transition element from the excited ($|e \rangle$)
to the virtual ($| v \rangle$) states.
For Majorana neutrino, its field 
$\nu_{i} \sim u a_i + v a^\dagger_i$ contains only a single 
type of creation/annihilation operator $a_i$. 
Both $\overline \nu$ and $\nu$ of the neutrino
current $j^{(\nu)}_\mu = \overline \nu_i \gamma^\mu (1 - 
\gamma_5) \nu_j$ can contract with 
any of the two final-state neutrinos.
In addition to the usual contractions $\langle \nu_i 
(p_{\nu_i})|\overline \nu 
\sim \overline u(p_{\nu_i})$ and 
$\langle \overline \nu_j (\overline 
p_{\nu_j}) |\nu \sim v(\overline 
p_{\nu_j})$ that contributes to the Dirac matrix
element in \geqn{eq:Mw-Dirac}, another two contractions 
are possible with $\langle \nu_i (p_{\nu_i})|\nu \sim 
v(p_{\nu_i})$ and $\langle \overline 
\nu_j (\overline p_{\nu_j}) |\overline 
\nu \sim \bar u(\overline p_{\nu_j})$
that appears for Majorana neutrinos
\cite{Dinh:2012qb}. A relative sign
in between comes from switching the fermion
fields (operators). Putting everything together, 
the matrix element for Majorana neutrinos is,
\begin{eqnarray}
  \mathcal M_W^{(M)}
=
  \sqrt{2} G_{\rm F} 
  (0, {\boldsymbol s}_{ve})^\mu 
  \left[ 
  a_{ij} 
  \bar u
  (p_{\nu_i}) 
    \gamma_\mu P_L 
  v (p_{\overline{\nu}_j})
- a_{ij}^* 
  \bar u(p_{\overline \nu_j})
    \gamma_\mu P_L 
  v (p_{\nu_i})
  \right],
\label{eq:Mw-Maj}
\end{eqnarray}
where we have implemented the property that
the axial vector coupling matrix is Hermitian,
$a_{ji} = a^*_{ij}$, according to its definition
below \geqn{eq:xsec_nueTOnue}.

In the presence of neutrino EM interactions, the neutrino 
pair emission receives extra contribution. The vertex
in \geqn{eq:nuEMcurrent} connects with the electron 
QED interaction ($e \bar \psi_e \gamma^\mu \psi_e$) via 
photon mediatior. The resulting effective Lagrangian is
\begin{align}
 \mathcal L_{EM} 
=
- e \bar \psi_e \gamma^\mu \psi_e
  \widetilde j_\mu^{(\nu EM)},
\end{align}
where the neutrino current in the momentum space is
$j_\mu^{(\nu EM)} \equiv \overline u(p_{\nu_i})
\Lambda^{ij}_\mu(q) v(p_{\bar \nu j}) / q^2$
which contributes to the vector current
$(J_V)_\mu \rightarrow j_\mu^{(\nu EM)}$ as
defined in \gapp{sec:Hemission},
\begin{align}
  H_{\nu EM}
& = 
- i e {\boldsymbol j}^{(\nu EM)} \cdot [\hat{\boldsymbol  x}, H_0]
+ \frac{e}{2m_e} \boldsymbol\sigma \cdot 
  (\boldsymbol\nabla \times {\boldsymbol j}^{(\nu EM)}),
\label{eq:HnuEM}
\end{align}
in a similar way as photon in \geqn{eq:Hgamma}.
With the neutrino pair emission being practically
selected to be an M1 transition, only the second
term can survive although suppressed by the electron mass.
For the purpose of probing the neutrino EM properties,
it is much more advantageous to select E1 transition
since the first term in \geqn{eq:HnuEM} 
is proportional to the dipole operator and 
typically $\alpha^{-1} \sim 100$ faster than the M1 
transition. Moreover, selecting the E1 type transition
can also reduce the neutrino pair emission background from
the SM weak $W/Z$ interactions by 2 orders. So an ideal RENP 
experiment for probing neutrino EM form factors should use
the E1$\times$E1 transitions. However, we still
adopt the E1$\times$M1 type configuration to make
a conservative illustration.

The corresponding matrix
element for the M1 transition is,
\begin{eqnarray}
    \langle v | H_{\nu EM}|e\rangle
\equiv
  \mathcal M_{\nu EM} e^{-i(p_\nu+p_{\bar \nu})\cdot x},
\quad {\rm with} \quad 
  \mathcal M_{\nu EM} 
=
  \frac {2 \mu_B}{q^2}
  (0, \boldsymbol s_{vg} \times \boldsymbol q)^\mu
~ \overline u(p_{\nu_i}) \Lambda_\mu^{ij}(q) v(p_{\bar \nu_j}).
\label{eq:Mnu_EM}
\end{eqnarray}
If E1$\times$E1 can be
experimentally implemented, the signal strength can be
further enhanced by 4 orders while the background is
suppressed by 4 orders.

\subsection{Kinematics and Phase Space}

To enhance the decay rate, a laser stimulation is needed.
As a consequence, the photon momentum magnitude
$|{\boldsymbol p}_\gamma| = \omega < \omega '$
and direction ($\hat {\boldsymbol p}_\gamma$) are fixed
by the incoming laser.
While the neutrino pair carries the energy dictated
by the atomic level difference $E_{eg} \equiv E_e - E_g$
and the photon energy $\omega$, the momentum conservation
requires the neutrino pair momentum to be opposite of
the emitted photon,
\begin{equation}
  q
\equiv
  (E_{eg} - \omega, - \omega \hat{\boldsymbol p}_\gamma)
= p_{\nu_i} + p_{\bar \nu_j}.
\label{eq:neutrino_pair_momentum}
\end{equation}
The equation above imposes a relation 
between the trigger laser frequency and 
the neutrino pair invariant mass,
$m^2_{\bar \nu \nu} \equiv (p_{\nu_i} + p_{\bar \nu_j})^2 = q^2$,
\begin{eqnarray}
  \omega 
=
  \frac{E_{eg}} 2
- \frac{m^2_{ \nu \bar\nu}}{2 E_{eg}}.
\end{eqnarray}
Because of the minus sign in front of $m^2_{\nu\bar \nu}$,
a maximum frequency threshold occurs at the minimum
neutrino pair invariant mass $\left. m^2_{ \nu\bar\nu} \right|_{\rm min} = (m_i+m_j)^2$ 
\cite{Song:2015xaa,Zhang:2016lqp,Dinh:2012qb},
\begin{eqnarray}
    \omega 
\leq 
  \omega_{ij}^{\rm max} 
\equiv 
  \frac{ E_{eg}} 2
- \frac{(m_i+m_j)^2}{2 E_{eg}},
  \label{eq:wij_max}
\end{eqnarray}
with $m_i$ ($m_j$) being the (anti-)neutrino mass.

If the neutrino pair is replaced by
a photon, the two back-to-back stimulating lasers have the same
frequency, $\omega = \omega' = E_{eg}/ 2$
with vanishing $m^2_{\nu \bar \nu}$.
Since the neutrino pair emission 
only occurs for $\omega < \omega'$, the two-photon
emission background $|e\rangle \rightarrow |g\rangle 
+ 2 \gamma$ can be suppressed by choosing
the stimulating lasers with two different
frequencies. It is practically possible to use
the E1$\times$E1 configuration to enhance the
signal rate from the neutrino EM properties
and at the same time suppressing the two-photon
background. Although being higher-order in $\alpha$, 
other QED backgrounds exists. For example, replacing the neutrino 
pair by a photon pair or any final state with more photons 
($|e\rangle \rightarrow |g\rangle + n\gamma$,  $n \geq 3$) 
produces a higher number of events than the RENP process
\cite{Yoshimura:2015fna}. Extra measures are needed to 
reduce the higher order QED backgrounds to a reasonable 
level. The use of crystal waveguides is a possible solution 
to achieve an $n$-photon background free experiment 
\cite{Tanaka:2016wir,Tanaka:2019blr}. In our work we 
consider that all QED backgrounds can be removed.

The three-body final-state
phase space reduces to the two-body
one for the two final-state neutrinos \cite{Fukumi:2012rn} 
and the differential decay rate is
\cite{Fukumi:2012rn,Dinh:2012qb},
\begin{equation}
  d \Gamma_{ij} 
=
  (2\pi)^4 \delta^{(4)}(q - p_{\nu_i} -  p_{\bar{\nu}_j})
  \overline{|\mathcal M_{e\rightarrow g}|^2}
  \frac{d^3 {\boldsymbol p}_{\nu_i}}
  {(2 \pi)^3 2 E_{\nu_i}}
  \frac{d^3 {\boldsymbol p}_{\nu_j}}
  {(2 \pi)^3 2 E_{\bar \nu_i}},
\label{eq:dif_decay_rate}
\end{equation}
where the total matrix element $\mathcal 
M_{e\rightarrow g}$ is obtained from the
second-order perturbation theory. The full element
$\mathcal M_{e\rightarrow g}$ is a combination of 
$\mathcal M_D$ in \geqn{eq:Me_dipole} and 
the neutrino pair emission matrix element 
$\mathcal M_{\bar \nu \nu}$ \cite{Fukumi:2012rn},
\begin{equation}
  \mathcal M_{e\rightarrow g}
=
 n_a \frac{\mathcal M_D \mathcal{M}_{\bar \nu \nu}}
 {E_{vg}-\omega},
\end{equation}
with energy difference $E_{vg} \equiv E_v - E_g$
and the atom number density $n_a$. The neutrino
emission matrix contains the two terms, 
$\mathcal{M}_{\bar \nu \nu} 
\equiv \mathcal M_W + \mathcal M_{\nu EM}$,
from the EW and EM contributions in \geqn{eq:Mw-Dirac}
and \geqn{eq:Mnu_EM}, respectively.

\subsection{Magnetic Moment and Electric Dipole Contributions} 
\label{sec:RENP_mu_e}

When only neutrino magnetic and electric  
moments are present, \geqn{eq:EMvertex}
simplifies and $\mathcal M_{\nu EM}$ becomes,
\begin{eqnarray}
  \mathcal M_{\mu,\epsilon}
\equiv
  \frac {2\mu_B}{q^2}
  (0, {\boldsymbol s}_{vg} \times  {\boldsymbol q})^\mu
~\overline u(p_{\nu_i}) 
  \left[ 
    (\mu_\nu)_{ij}
    \sigma_{\mu\beta} q^\beta
  +(\epsilon_\nu)_{ij}
    \sigma_{\mu\beta} q^\beta \gamma_5
  \right]
  v(p_{\nu_j}) 
  \,.
  \label{eq:MEM_eps_mu}
\end{eqnarray}
The integration $d^3 {\boldsymbol p}_{\bar\nu_j}$
in \geqn{eq:dif_decay_rate} removes the three-dimensional
$\delta$ function for the momentum conservation
while the remaining $\delta$ function for the energy
conservation adapts the angular part of
$d^3 {\boldsymbol p}_{\nu_i}$  to correlate the
opening angle $\theta_0$ between the neutrino and photon
momentums,
\begin{eqnarray}
 \cos \theta_0
\equiv 
  \frac{
    q^2 
  - \Delta m^2_{ji}
  - 2E_{\nu_i}(E_{eg} - \omega)
   }{2 \omega |{\boldsymbol p}_{\nu_i}|},
\label{eq:cos_thet0_def}
\end{eqnarray}
where $\Delta m_{ij}^2 \equiv m_i^2 - m_j^2$
and $q^2 = E_{eg}^2 - 2E_{eg} \omega$ are
obtained from \geqn{eq:neutrino_pair_momentum}. 
The geometric limits $|\cos  \theta_0| \leq 1$
imply an allowed neutrino energy range,
\begin{equation}
  \overline E - \frac{\omega \Delta_{ij}(\omega)} 2
< E_{\nu_i} <
  \overline E + \frac{\omega \Delta_{ij}(\omega)} 2,
\label{eq:EnuLimits}
\end{equation}
where,
\begin{eqnarray}
  \overline E
\equiv 
   \frac {(E_{eg}-\omega) [q^2 -  \Delta m^2_{ji}]}{2 q^2} ,
\quad {\rm and} \quad 
  \Delta_{ij}(\omega)
\equiv
  \frac {\sqrt{[q^2-(m_i+m_j)^2] [q^2-(m_i-m_j)^2]}}{q^2}
    \,.
\label{eq:barE_Dij}
\end{eqnarray}
Putting everything together, the differential 
decay rate in \geqn{eq:dif_decay_rate} becomes 
a function of the neutrino energy $E_{\nu_i}$,
\begin{eqnarray}
  \frac{d\Gamma_{ij}}{d E_{\nu_i}} 
=
  \frac {n^2_a}{8 \pi \omega}
  \frac {|{\boldsymbol d}_{gv} \cdot {\boldsymbol E}|^2}
  {(E_{vg} - \omega)^2} 
  \left. 
    \overline{|\mathcal M_{\bar \nu \nu}|^2} 
  \right|_{\theta = \theta_0}.
\label{eq:dGammadEnu}
\end{eqnarray}

The spin-average square of $\mathcal M_{\bar \nu 
\nu}$ has three parts. First, the weak-only component
\cite{Dinh:2012qb,Song:2015xaa,Zhang:2016lqp,
Ge:2021lur} is,
\begin{eqnarray}
  \overline{|\mathcal M_W^{ij}|^2} 
=
  \frac 8 3(2 J_v + 1) |a_{ij}|^2 G_{\rm F}^2 C_{ev}
\Bigl\{
  2 ({\boldsymbol p}_{\nu_i}\cdot {\boldsymbol p}_{\bar \nu_j})
+ 3 (p_{\nu_i} \cdot p_{\bar \nu_j})
- 3 \delta_M {\rm Re}[a_{ij}^2] m_i m_j
\Bigr\},
\label{eq:MW_average}
\end{eqnarray}
where the quantity $\delta_M = 0$ for Dirac neutrinos 
and is $\delta_M = 1$ for Majorana neutrinos.
The extra term $\propto 3 \delta_M {\rm Re}[a_{ij}^2] m_i m_j$ that appears only for the Majorana neutrino case effectively changes the distribution of emitted photons and can be used to distinguish from the Dirac case and even the Majorana phases \cite{Fukumi:2012rn,Dinh:2012qb}.
The averaging is performed over the initial- and 
final-state spin states. For convenience, the
spin operators have been
eliminated using the spherical symmetry property of 
atomic states \cite{Dinh:2012qb},
\begin{equation}
  \frac{1}{(2 J_e + 1)}\sum_{\rm spins}
  {\boldsymbol s}^i_{vg}({\boldsymbol s}^j_{vg})^*
=
  (2 J_v + 1) 
  \frac{C_{ev}}{3} 
  \delta^{ij},
  \label{eq:Sij}
\end{equation}
where $J_e$ and $J_v$ are the total spins of the excited
($|e\rangle$) and virual ($|v\rangle$) states,
respectively. The coefficient $C_{ev}$ is a numerical 
factor that is different for each atom.
For both Xe and Yb, $(2 J_v +1)C_{ev} = 2$ \cite{Song:2015xaa}.

The interference term is a combination of the
weak and EM components, $2{\rm Re}[\mathcal M_W 
\mathcal M_{\mu,\epsilon}^*]$,
\begin{align}
\nonumber 
  \overline{\mathcal M_W^{(D)} \mathcal M_{\mu,\epsilon}^*}
& =
-
  \frac{
    2\sqrt{2} G_{\rm F}a_{ij}\mu_B
  }{q^2}
  \frac{1}{2(2 J_e + 1)}
  \sum_{\rm spins}
  (0, {\boldsymbol  s}_{ev})^\mu 
  (0, {\boldsymbol  s}_{vg} \times {\boldsymbol q} )^\nu
\\
& \times
  {\rm Tr} 
  \left\{
    \gamma_\mu P_L 
    (\slashed p_{\overline \nu_j} - m_j)
 \left[ 
   (\mu_\nu)_{ij}^*\sigma_{\nu\beta}q^\beta
 + (\epsilon_\nu)_{ij}^*
   \sigma_{\nu\beta} q^\beta \gamma_5
 \right]
    (\slashed p_{\nu_i} + m_i)      
\right\},
\end{align}
for Dirac neutrinos. After applying the spin-operator
identity in \geqn{eq:Sij}, the interference term becomes,
\begin{eqnarray}
\hspace{-3mm}
  \overline{ \mathcal M_W^{(D)} \mathcal M_{\mu,\epsilon}^* }
=
  \frac{
    8 C_{ev} (2 J_v + 1) 
    \sqrt 2 G_{\rm F}\mu_B a_{ij}
  }{3}
  \left[ 
    (\mu_\nu)^*_{ij} (m_i + m_j)
  +(\epsilon_\nu)^*_{ij}(m_i - m_j)
  \right]
  (\overline E - E_{\nu_i}),
\quad
\label{eq:Iint_MWMEM}
\end{eqnarray}
where $\overline E$ is defined in \geqn{eq:barE_Dij}. 
The only difference between the neutrino
magnetic moment and electric dipole is a minus sign in 
front of $m_j$ from the presence of $\gamma_5$ in
the $(\epsilon_\nu)_{ij} \sigma^{\mu \nu}\gamma_5$
vertex. While $\overline{ \mathcal M_W \mathcal 
M_{\mu,\epsilon}^*}$ is proportional to
$\overline E - E_{\nu_i}$, the range of $E_{\nu_i}$ in 
\geqn{eq:EnuLimits} is anti-symmetric around $\overline E$. 
Consequently, the interference contribution to the
total rate $\Gamma$ is zero. Similar cancellation occurs
for Majorana neutrinos. Although such interference term
can appear in the differential rate $d \Gamma / d E_{\nu_i}$,
the neutrino energy cannot
be practically reconstructed due to its illusiveness.

The new physics contribution to the total rate $\Gamma$
comes from the EM-only term, $|\mathcal M_{\mu,\epsilon}|^2$,
\begin{align}
  \overline{|\mathcal M_{\mu,\epsilon}|^2} 
& =
  \frac{1}{2 (2J_e + 1)}
  \sum_{\rm spins} 
  \frac{4\mu_B^2}{q^4}
  (0, {\boldsymbol s}_{vg}\times {\boldsymbol q})^\mu
  (0, {\boldsymbol s}_{vg}^*\times {\boldsymbol q})^\nu
\\
& \nonumber 
  {\rm Tr}
  \left\{ 
    \left[ 
      (\mu_\nu)_{ij}
      \sigma_{\mu\alpha} q^\alpha
    +(\epsilon_\nu)_{ij}
      \sigma_{\mu\alpha} q^\alpha\gamma_5
    \right]
    (\slashed p_{\nu_i} + m_i)
    \left[ 
      (\mu_\nu)_{ij}^*
      \sigma_{\mu\beta} q^\beta
    + (\epsilon_\nu)_{ij}^*
      \sigma_{\mu\beta} q^\beta\gamma_5
    \right]
  (\slashed p_{\nu_j} - m_j)
  \right\}.
\end{align}
Using the relation in \geqn{eq:Sij} and further
noticing that there is no interference between 
the magnetic ($\mu_\nu$) and electric ($\epsilon_\nu$)
moment contributions due to 
the presence of a $\gamma_5$ matrix in the 
electric dipole interaction, the matrix 
element squared becomes,
\begin{eqnarray}
  \overline{ |\mathcal M_{^\mu_\epsilon}^{ij}|^2 }
=
  C_{ev}
  (2J_v + 1)
  \frac{8 \mu_B^2\omega^2}{3q^4}
  \left[
    q^2 (m_i \pm m_j)^2 
  - (\Delta m_{ji}^2)^2
  + 2 q^2 |{\boldsymbol p}_i|^2 \sin^2 \theta 
  \right]
\begin{cases}
  |(\mu_\nu)_{ij}|^2, \\[5pt]
  |(\epsilon_\nu)_{ij}|^2.
\end{cases}
\end{eqnarray}
Similar to the Weak/EM interference term, the difference 
between the electric and magnetic moment
contributions is a minus sign in front of $m_j$.

Putting this result back into the differential
decay rate in \geqn{eq:dGammadEnu}, together 
with the definition of $\theta_0$ in 
\geqn{eq:cos_thet0_def} and the integration 
range in \geqn{eq:EnuLimits}, the eletromagnetic 
total decay rate becomes \cite{Ge:2022cib},
\begin{eqnarray}
  \Gamma_{^\mu_\epsilon} 
=
  \Gamma_0
\sum_{ij} 
  \frac{\omega^2 \mu^2_B}{G^2_F}
  \frac{
    \Delta_{ij}
    \Theta(\omega - \omega_{ij}^{\rm max})
  }{9(E_{vg} - \omega)^2}
  \Biggl[
    1
   +\frac{(m_i \pm m_j)^2 \pm 4m_i m_j}{q^2}
  -2\left(\frac{\Delta m_{ji}^2}{q^2}\right)^2
\Biggl]
\begin{cases}
    |(\mu_\nu)_{ij}|^2, 
\\[5pt] 
    |(\epsilon_\nu)_{ij}|^2,
\end{cases}
\label{eq:Gamma_mu_eps}
\end{eqnarray}
where the Heaviside $\Theta$-functon 
defines the $\omega_{ij}^{\rm max}$ 
thresholds in \geqn{eq:wij_max}. The 
rate can be charactrized by a benchmark
value $\Gamma_0$ \cite{Song:2015xaa,
Zhang:2016lqp},
\begin{eqnarray}
  \Gamma_0 
\equiv 
  (2J_v + 1)\frac{
    n_a^2 C_{ev} G_{\rm F}^2 \langle|{\boldsymbol d}_{gv}\cdot 
    {\boldsymbol E}|^2\rangle }{\pi }
\approx
  0.002{\rm s}^{-1}
  \left(\frac{V}{10^2 {\rm cm}^3}\right)
  \frac{n_a^2 n_\gamma }{\left(10^{21}{\rm cm}^{-3}\right)^3}
  \eta(t),
  \label{eq:Gamma0_def}
\end{eqnarray}
with $\eta$ describing the fraction of the 
target volume $V$ that is macroscopically coherent.
The spectral function is then defined as the relative
size, $\mathcal I_{\mu,\epsilon} \equiv 
\Gamma_{\mu,\epsilon}/\Gamma_0$. 

\begin{figure}[t]
  \centering
  \includegraphics[scale = 0.52]{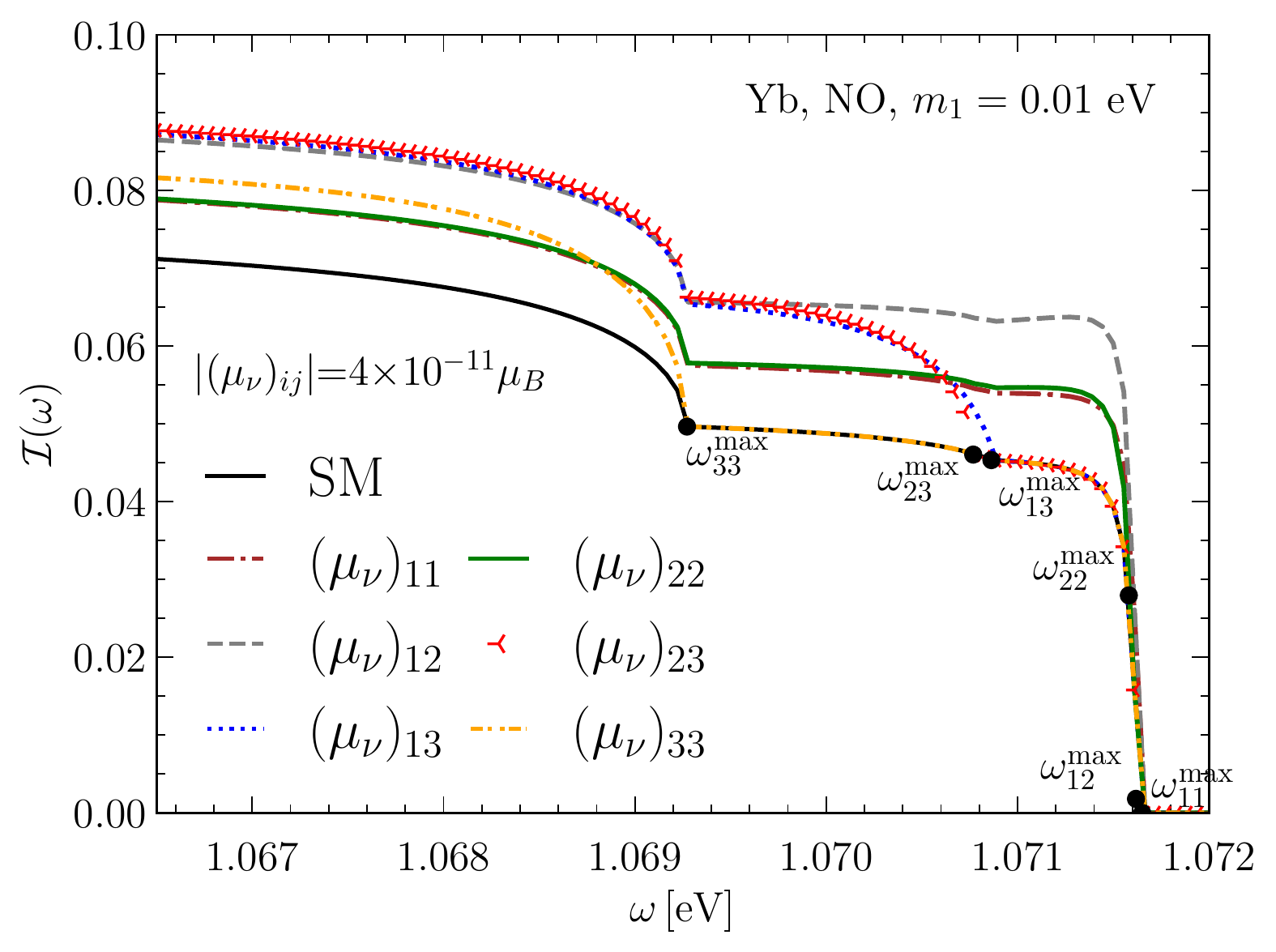}
  \includegraphics[scale = 0.52]{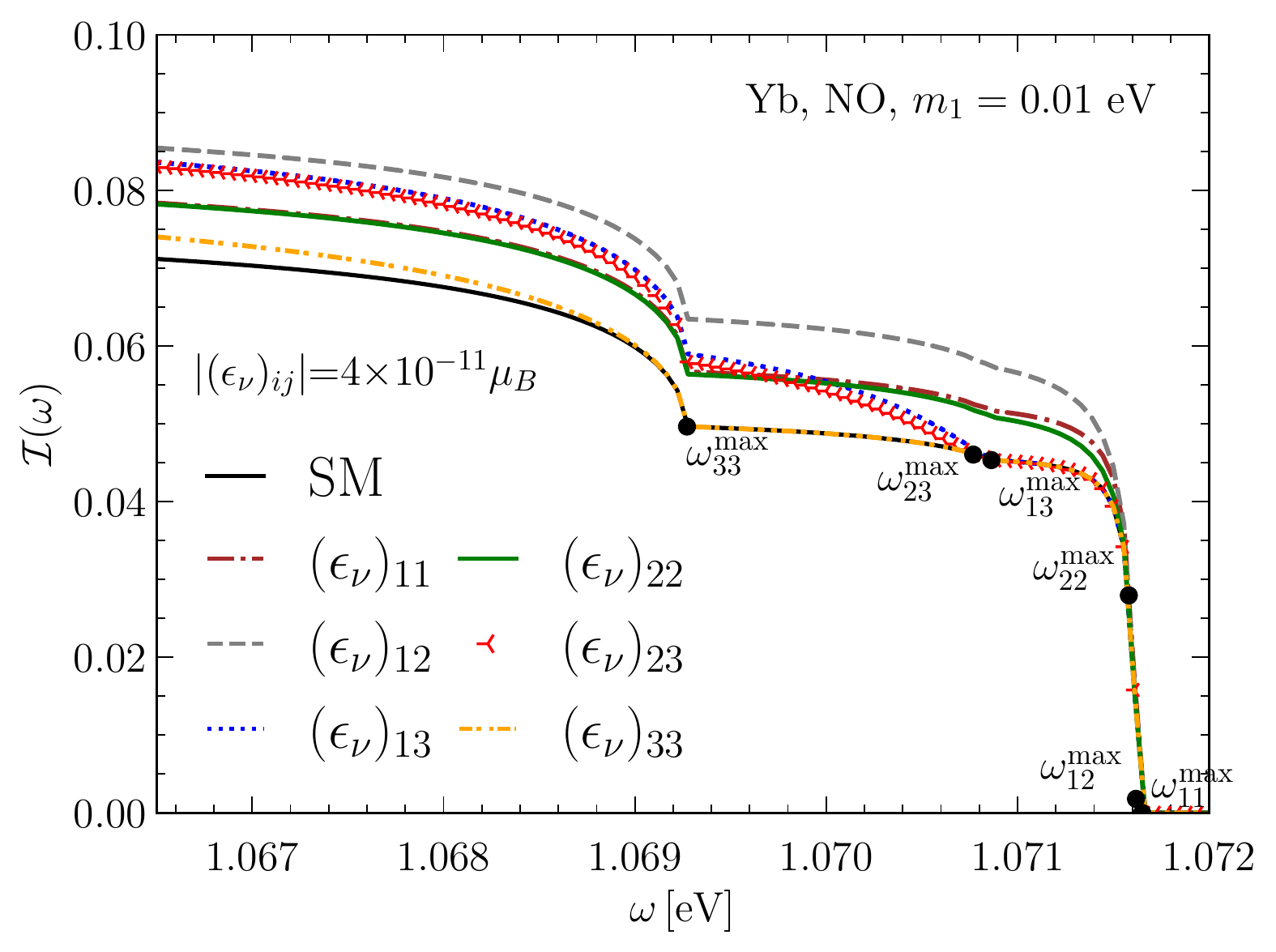}
  \includegraphics[scale = 0.52]{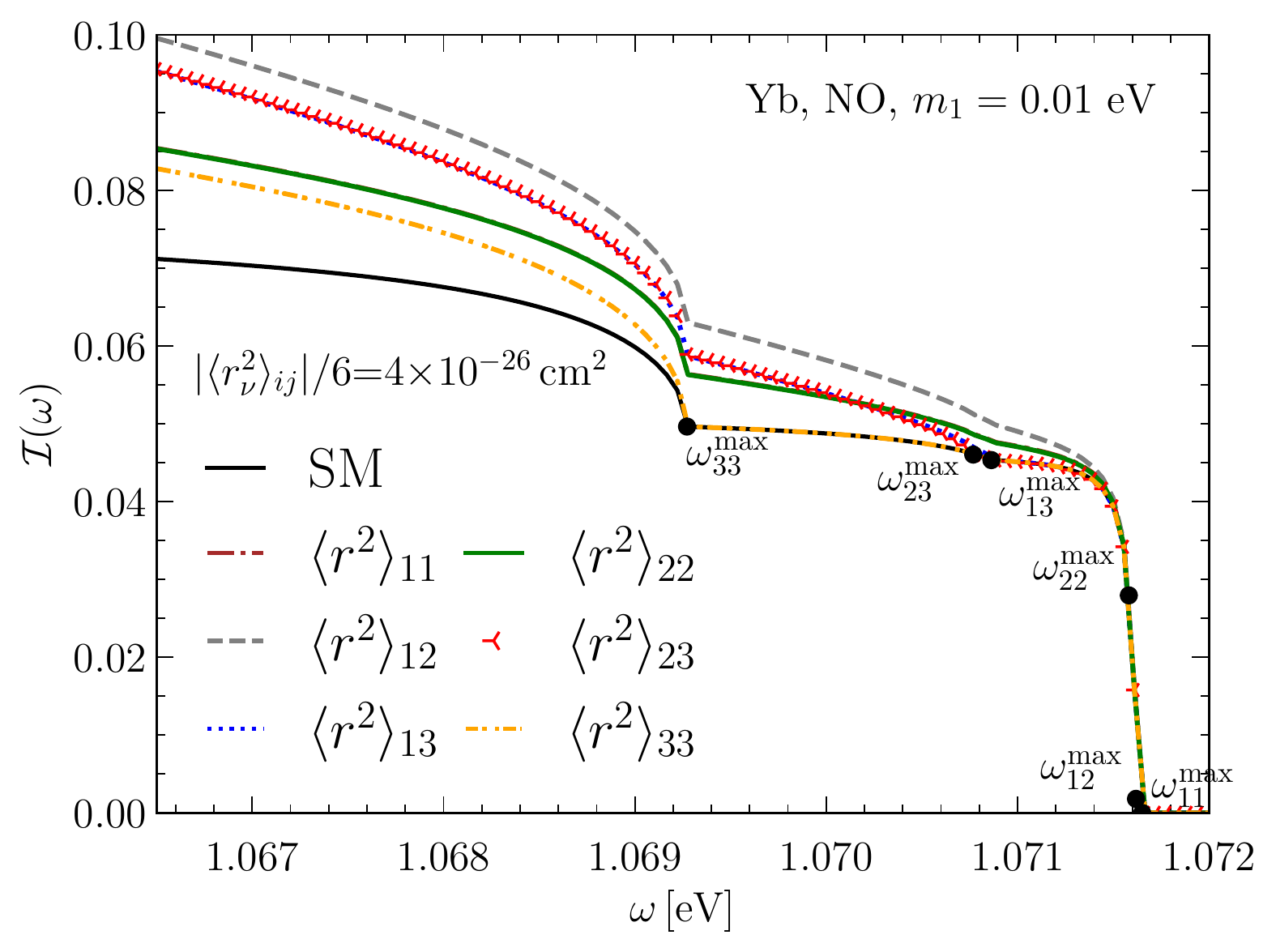}
  \includegraphics[scale = 0.52]{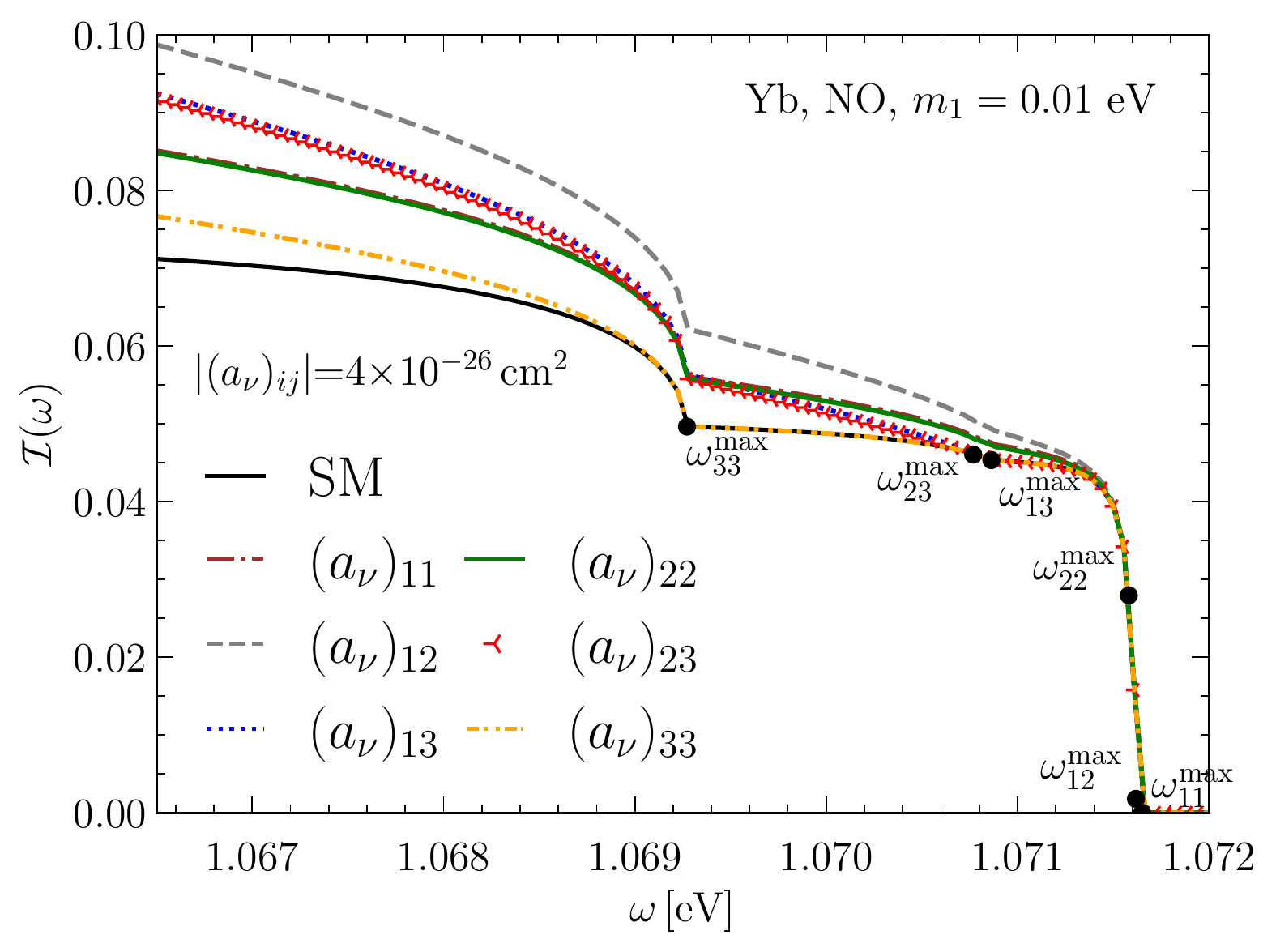}
\caption{
The spectral function $\mathcal I(\omega)$
for the Yb atom RENP processes with weak (SM) and neutrino
EM interactions as a function of the trigger laser frequency
$\omega$ for the normal ordering (NO) of neutrino masses
and the lightest neutrino mass $m_1 = 0.01$\,eV.}
\label{fig:I_mu_eps}
\end{figure}

Comparing the result in \geqn{eq:Gamma_mu_eps} 
with the SM-only case
\cite{Dinh:2012qb,Song:2015xaa,Zhang:2016lqp,
Ge:2021lur},
\begin{align}
  \Gamma_W
& =
  \Gamma_0 
\sum_{ij} 
  \frac{
    \Delta_{ij}
    \Theta(\omega - \omega_{ij}^{\rm max})
  }{3(E_{vg} - \omega)^2}
\Biggl\{
    |a_{ij}|^2 
\Bigl[ 
      q^2
      -\frac{1}{2}(m_i^2+m_j^2)
      +\frac{\omega^2}{2}
    \left(1-\frac{1}{3}\Delta_{ij}^2\right)
\nonumber
\\
& \hspace{50mm}
-\frac{
      (E_{eg}-\omega)^2(\Delta m_{ij}^2)^2
     }{2q^4}
\Bigr]
  -3\delta_M {\rm Re}[a_{ij}^2] m_i m_j
\Biggr\},  
\label{eq:Gamma_W}
\end{align}
the relative size $\Gamma_\mu/\Gamma_W \sim 
\omega^2 \mu_B^2|(\mu_\nu)_{ij}|^2 /
G_F^2  q^2 \sim |(\mu_\nu)_{ij}|^2 /
[10^{-11}\mu_B]^2$ scales with $\mu^2_\nu$
for the magnetic moment and similarly for
its electric dipole counterpart. While the SM
contribution is mediated by the heavy $W/Z$ bosons,
the neutrino magnetic and electric moment contributions
are inversely proportional to $1/q^4$.
With typical $q \sim \mathcal O$(eV) for atomic processes,
the SM contribution is suppressed by a factor of
$(q^2/m_W^2)^2 \sim 10^{-44}$.
However the presence of the electron magnetic moment 
suppresses the EM contribution by $\omega/\mu_B\sim 10^{-6}\,$eV.
All in all, a $|(\mu_\nu)_{ij}| \sim 10^{-11} \mu_B$
provides similar contribution to that of the SM.

The total spectral function,
$\mathcal I \equiv \mathcal I_W 
+ \mathcal I_{\mu, \epsilon}$, is presented in the 
upper left (right) panels of \gfig{fig:I_mu_eps} with
the representative value of $|(\mu_\nu)_{ij}|^2$ and $|(\epsilon_\nu)_{ij}|^2$ being 
$4\times10^{-11}\,\mu_B$. In both cases, the 
curves significantly differ from the SM contribution 
(black). The kinks in the curves represents the 
location of all six frequency thresholds in 
\geqn{eq:wij_max}. In all curves, the magnetic 
moment contribution is slightly larger than the 
electric dipole one due to the difference in the 
sign in $m_j$. The difference in the shape of 
$\mathcal I$ over the frequency range can be used to 
experimentally distinguish both contributions.

\subsection{Charge Radius and Anapole Contributions} 
\label{sec:RENP_a_r}

The charge radius and anapole always appear together
with similar Lorentz structure,
\begin{eqnarray}\label{eq:Mar}
  \mathcal M_{r,a}
=
 2\mu_{\rm B}
  (0, {\boldsymbol  s}_{vg} \times {\boldsymbol q})_\mu
  \overline u(p_{\nu_i}) 
  \left(
    \gamma^\mu 
  -\frac{q^\mu \slashed q}{q^2} 
  \right) 
  \left(
    \frac{\langle r^2_\nu \rangle_{ij}}{6} 
  + (a_\nu)_{ij}  
    \gamma_5 
  \right) 
  v(p_{\nu_j}).
\end{eqnarray}
The total matrix element is then
$\mathcal M_{\nu\bar\nu} = \mathcal M_W + \mathcal M_{r,a}$.
Similarly to the previous section, the spin averaged matrix
element squared contains three terms, the SM-only term,
the interference term between SM and EM contributions,
as well as the pure charge radius and anapole terms.
The interference between $M_W^{(D)}$ in \geqn{eq:Mw-Dirac}
and $\mathcal M_{r,a}$ is,
\begin{align}
\overline{
   \mathcal M_{W}^{(D)}\mathcal M_{r,a}^*
   }
= &
-\frac{2\sqrt{2} G_{\rm F}\mu_B a_{ij}}{2(J_e + 1)} 
  \sum_{spins}
  (0, \hat{\boldsymbol  s}_{vg})^\nu 
  (0, {\boldsymbol  s}_{vg}\times {\boldsymbol q})^\mu 
\nonumber
\\
& \times
  {\rm Tr}
  \Biggl\{
    \left(
      \gamma_\mu 
    - \frac{q_\mu \slashed q}{q^2} 
    \right) 
    \left[
      \frac{\langle r^2_\nu \rangle_{ij}^*}{6} 
    +(a_\nu)_{ij}^*  \gamma_5 
    \right]
    (\slashed p_j - m_j) 
    \gamma_\nu P_L 
    (\slashed p_i + m_i) 
  \Biggr\}.
\label{eq:interference_ana_dirac}
\end{align}
The sum over electron spins is performed using the 
spin relation in \geqn{eq:Sij} to obtain,
\begin{eqnarray}
  2 {\rm Re} 
  \left[
  \overline{
    \mathcal M_{W}^{(D)}
    \mathcal M_{r,a}^*
   }
  \right]
=
  \frac{
    16\sqrt{2}(2 J_v + 1) C_{ve} \mu_{\rm B} G_{\rm F}}
  {3}
  q^2{\rm Im}
  \left[
    a_{ij}
    \left(
      \frac{\langle r^2_\nu \rangle_{ij}^*}{6}
    -(a_\nu)_{ij}^*
    \right)   
   \right]
  \left( E_{\nu_i} - \overline E \right).
\end{eqnarray}
with $\overline E $ defined in \geqn{eq:barE_Dij}. 
As before, this interference term ${\rm Re} \left[\overline{\mathcal M_{W}^{
(D)} \mathcal M_{r,a}^*}\right] \propto \left(
E_{\nu_i} - \overline E\right)$ is an
odd function around $\overline E$ and hence does
not contribute to the total rate after integration
over $E_{\nu_i}$.

The only contribution from the neutrino charge radius
and anapole is the square of \geqn{eq:Mar},
\begin{align}
  \overline{|\mathcal M_{r,a}|^2}
= &
 4\mu_{\rm B}^2
 \frac{1}{2(2 J_e + 1)}
 \sum_{\rm spins}
 (0, {\boldsymbol  s}_{ev}\times{\boldsymbol q})^\mu
 (0, {\boldsymbol  s}_{ev}^*\times {\boldsymbol q})^\nu
{\rm Tr} 
\Biggl[
  \left(
    \gamma^\mu 
  - \frac{q^\mu \slashed q}{q^2} 
  \right) 
\\ & 
\left(
  \frac{\langle r^2_\nu \rangle_{ij}}{6} 
+ (a_\nu)_{ij} \gamma_5 
\right)   
(\slashed p_{\overline \nu_j} - m_j)
\left(
  \gamma^{\nu} 
- \frac{q^{\nu} \slashed q}{q^2} 
\right) 
\left(
  \frac{\langle r^2_\nu \rangle_{ij}}{6} 
 +(a_\nu)_{ij}^* \gamma_5
\right)   
(\slashed p_{\nu_i} + m_i)
\Biggr].
\nonumber
\end{align}
The presence of $\gamma_5$ in the anapole term 
ensures that there is no interference between 
$\langle r^2_\nu \rangle_{ij}$ and $(a_\nu)_{ij}$. 
After using \geqn{eq:Sij} to eliminate the spin
operators, the matrix element squared becomes,
\begin{eqnarray}
  \overline{|\mathcal M_{^r_a}|^2}
=
  \frac{4 (\mu_{\rm B}\omega)^2}
  {3}(2 J_v + 1) C_{ve}
\Biggl[
    \frac{q^2}2
  - \frac{1}{2}
    (m_i \mp m_j)^2
  + |{\boldsymbol p}_i|^2 \sin^2 \theta
\Biggr]
\begin{cases}
  \frac{|\langle r^2_\nu \rangle_{ij}|^2}{36},
\\[5pt]
  |(a_{\nu})_{ij}|^2.
\end{cases}
\end{eqnarray}
The only difference between the charge radius
and anapole terms is the minus sign 
in front of $m_j$, which comes from the
$\gamma_5$ matrix in the anapole interaction. Using the 
decay rate definition in \geqn{eq:dGammadEnu} and 
the kinematic limits in \geqn{eq:EnuLimits}, 
the total decay rate becomes,
\begin{eqnarray}
  \Gamma_{^r_a}
=
  \Gamma_0
  \sum_{ij}
  \frac{|\mu_{\rm B}|^2}{9G_F^2}
  \frac {
    \omega^2 q^2\Delta_{ij}
    \Theta(\omega - \omega_{ij}^{\rm max})
  }{(E_{vg} - \omega)^2}
  \left[1 - \frac{(m_i \mp m_j)^2}{q^2}\right]
  \left[1 - \frac{(m_i \pm m_j)^2}{4q^2}\right]
\begin{cases}
  \frac{|\langle r^2_\nu \rangle_{ij}|^2}{36},
\\[5pt]
  |(a_{\nu})_{ij}|^2,
\end{cases}
\label{eq:Gammaar}
\end{eqnarray}
where $\Gamma_0$ is defined in \geqn{eq:Gamma0_def} and
the step function $\Theta (\omega - \omega_{ij}^{\rm max})$ 
defines the frequency thresholds of 
\geqn{eq:wij_max}. The relative size 
between the SM contribution in \geqn{eq:Gamma_W} 
and the anapole term is
$\Gamma_a/\Gamma_W \sim \mu_B^2\omega^2
|(a_\nu)_{ij}|^2 / 3 G_F^2 \sim 
 \left( \omega / {\rm eV} \right)^2
\left( |(a_\nu)_{ij}| / 10^{-26}{\rm cm}^{2}
\right)^2$. With the expectation that the sensitivity
is projected for the signal to have roughly the
same event rate as the SM background, the
neutrino charge radius and anapole can be probed
at the level of $10^{-26}{\rm cm}^{2}$.
Notice that an experiment with 
higher trigger frequency $\omega$ provides larger  
sensitivity on the charge radius and anapole due
to the $\omega^2$ factor.

The total spectral function 
$\mathcal I \equiv \mathcal I_W 
+ \mathcal I_{r,a}$ is presented in the 
bottom left (right) panels of \gfig{fig:I_mu_eps} 
for the representative value of $|\langle r^2_\nu 
\rangle_{ij}|/6$ and $|(a_\nu)_{ij}|$ of 
$4\times10^{-26}$\,cm$^2$. In both cases, the 
curves significantly differ from the SM contribution 
specially for lower frequencies. In all curves, the 
charge radius contribution is slightly larger than 
the anapole one due to the difference in the sign in 
$m_j$, which modified $(m_i \pm m_j)^2/q^2$ in 
\geqn{eq:Gammaar}. Since $q^2 = E_{eg}^2 - 2 
E_{eg}\omega$ decreases with increasing frequency, the 
relative difference between the charge radius and 
anapole curves becomes larger for higher frequency.
In the next section, we show that the difference in the 
shape of $\mathcal I_r$ and $\mathcal I_a$ can be
used to experimentally distinguish them.

\section{RENP Sensitivity to Neutrino EM Properties}
\label{sec:RENPsensitivity} 

\begin{figure}[!t]
\centering
\includegraphics[width=0.8\textwidth]{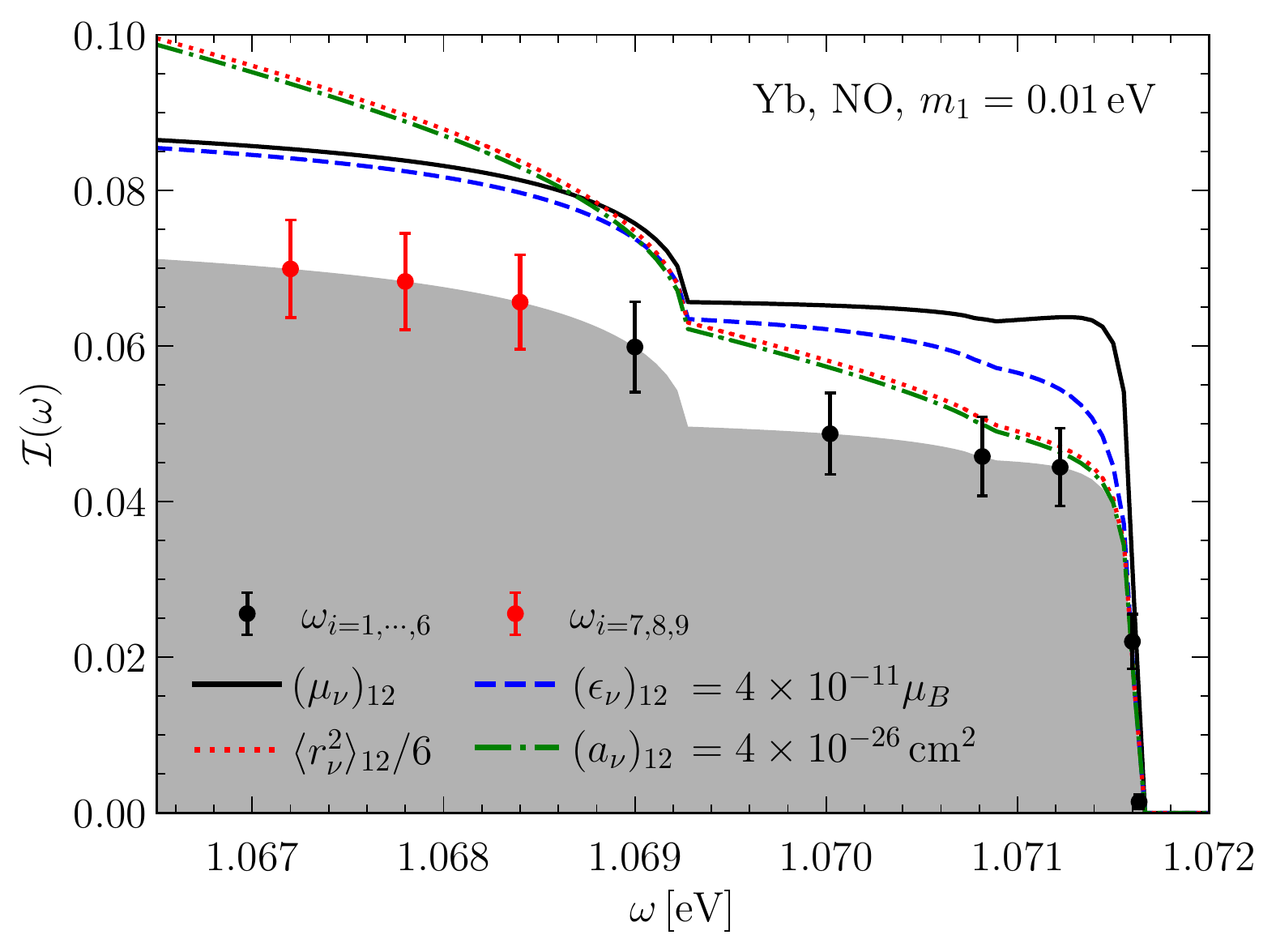}
\caption{
The spectral function $\mathcal I(\omega)$ for the Yb
atom RENP process considering SM background only (gray area),
the neutrino magnetic/electric moment
$(\mu_\nu)_{12} = 4 \times 10^{-11}\mu_B$ (black line),
$(\epsilon_\nu)_{12} = 4\times 10^{-11}\mu_B$
(blue dashed line),
$\langle r^2_\nu \rangle_{12} = 2.4 \times 10^{-25}$\,cm$^2$
(red dotted line), and
$(a_\nu)_{12} = 4 \times 10^{-26}$\,cm$^2$
(green dashed-dotted line) in the normal ordering (NO) case
with $m_1 = 0.01$\,eV. We also show the 6 frequencies
(black dots) needed to separate the matrix elements and
the other 3 frequencies (red dots) for identifying the form
factor type.
}
\label{fig:datapoints}
\end{figure}

The advantage of RENP is its capability of
separating the individual form factor elements
$f^{ij}_{M, E, Q, A}$ of \geqn{eq:EMvertex}
as first noticed in \cite{Ge:2022cib}. 
From the hermiticity condition, each form factor 
contains 6 elements that enter the spectral 
function as magnitudes squared $|f^{ij}_{M, E, Q, A}|^2$
while their phases do not appear. The kinematics
in \gsec{sec:RENP_mu_e} allows the photon energy
up to certain threshold frequency $\omega^{ij}_{\rm max}$
defined in \geqn{eq:wij_max} for each neutrino pair
$\nu_i \bar \nu_j$.
With the trigger laser having
a precision of $\delta \omega \sim 10^{-5}$\,eV 
\cite{Dinh:2012qb} which is much better than
$E_{eg}/2 - \omega_{ij}^{\rm max} 
\sim 10^{-5} - 10^{-3}$\,eV, it is possible to scan
the various thresholds $\omega_{ij}^{\rm max}$
with different combinations of neutrino mass
eigenvalues and the regions in between.
Scanning over 6 frequencies is
sufficient to separate individual matrix elements as
proposed in \cite{Ge:2022cib}. For illustration, we take
$\omega_i = (1.069$, 1.07, 1.0708, 1.0712, 1.0716, 
1.07164)\,eV shown as the black points in
\gfig{fig:datapoints}. 

Once nonzero deviation from the SM prediction is
observed, the experiment is also
capable of identifying which neutrino property
($\mu_\nu$, $\epsilon_\nu$, $\langle r^2_\nu\rangle$,
or $a_\nu$) is contributing. If one element
with specific indices $i$ and $j$ is experimentally
observed to be nonzero, the spectrum function below
the corresponding $\omega^{\rm max}_{ij}$ is affected
and can be scanned to identify which type that element
is. At least another
3 frequency points are necessary to reconstruct
the photon spectrum and then identify the relevant
neutrino property. Since the spectrum function
has larger value for smaller $\omega$, we choose the 
three frequencies $\omega_i = (1.0672$, 1.0678, 1.0684)\,eV
shown as red points in \gfig{fig:datapoints} to
increase event rate. Another reason of choosing
lower frequencies bellow all $\omega^{\rm max}_{ij}$
is that all elements can contribute to this region.
However, if the nonzero element has larger
$\omega^{\rm max}_{ij}$, it is also advantageous
to add more frequency points. Both the low and
high frequency regions are sensitive to distinguishing
$\langle r^2_\nu\rangle_{12}$ and $(a_\nu)_{12}$
from $(\mu_\nu)_{12}$ and $(\epsilon_\nu)_{12}$
with apparent difference. The difference between
$(\mu_\nu)_{12}$ and $(\epsilon_\nu)_{12}$ is also
larger at the high frequency region.

\begin{figure}[!t]
\centering
\includegraphics[width=0.49\textwidth]{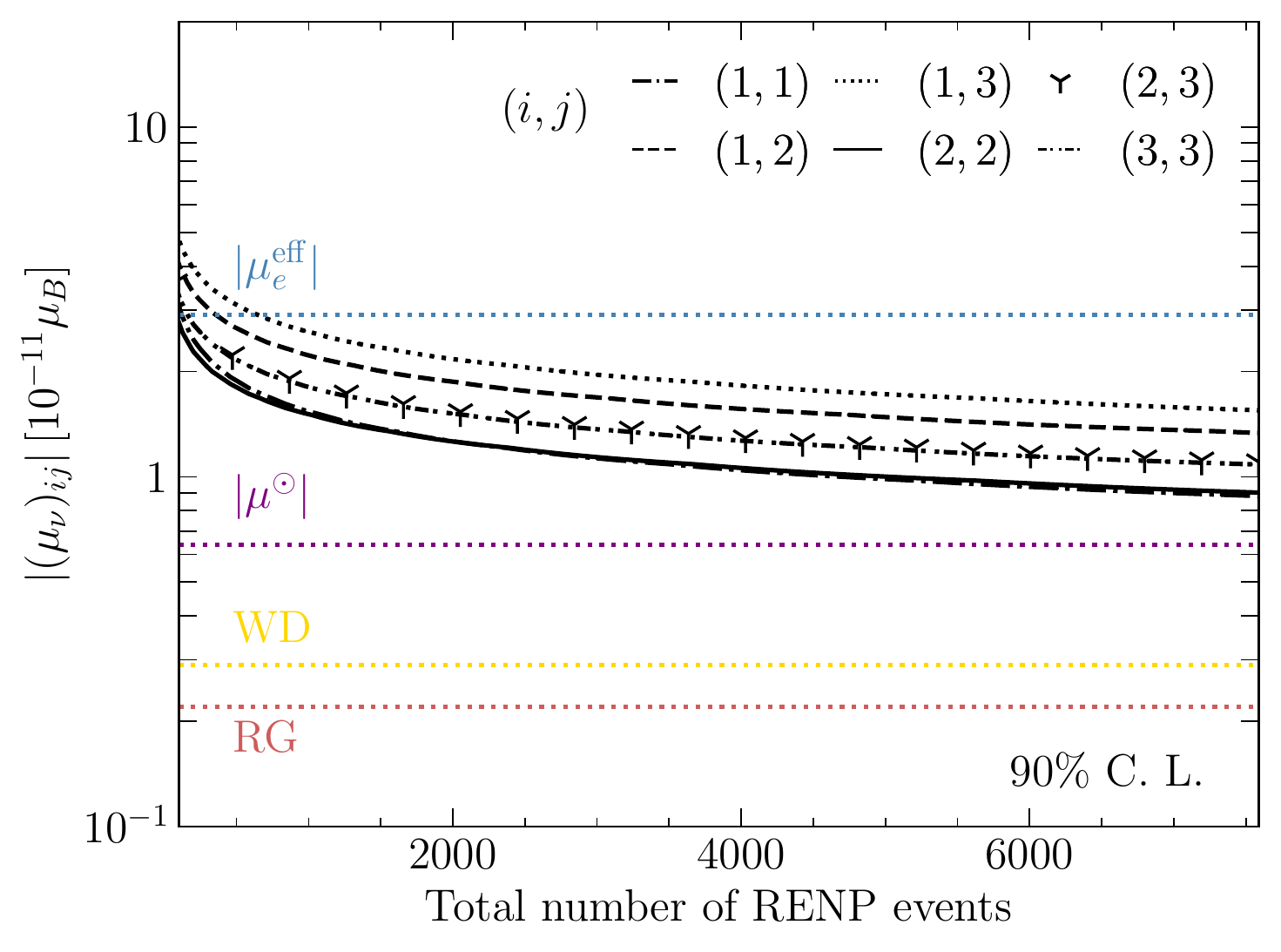}
\includegraphics[width=0.49\textwidth]{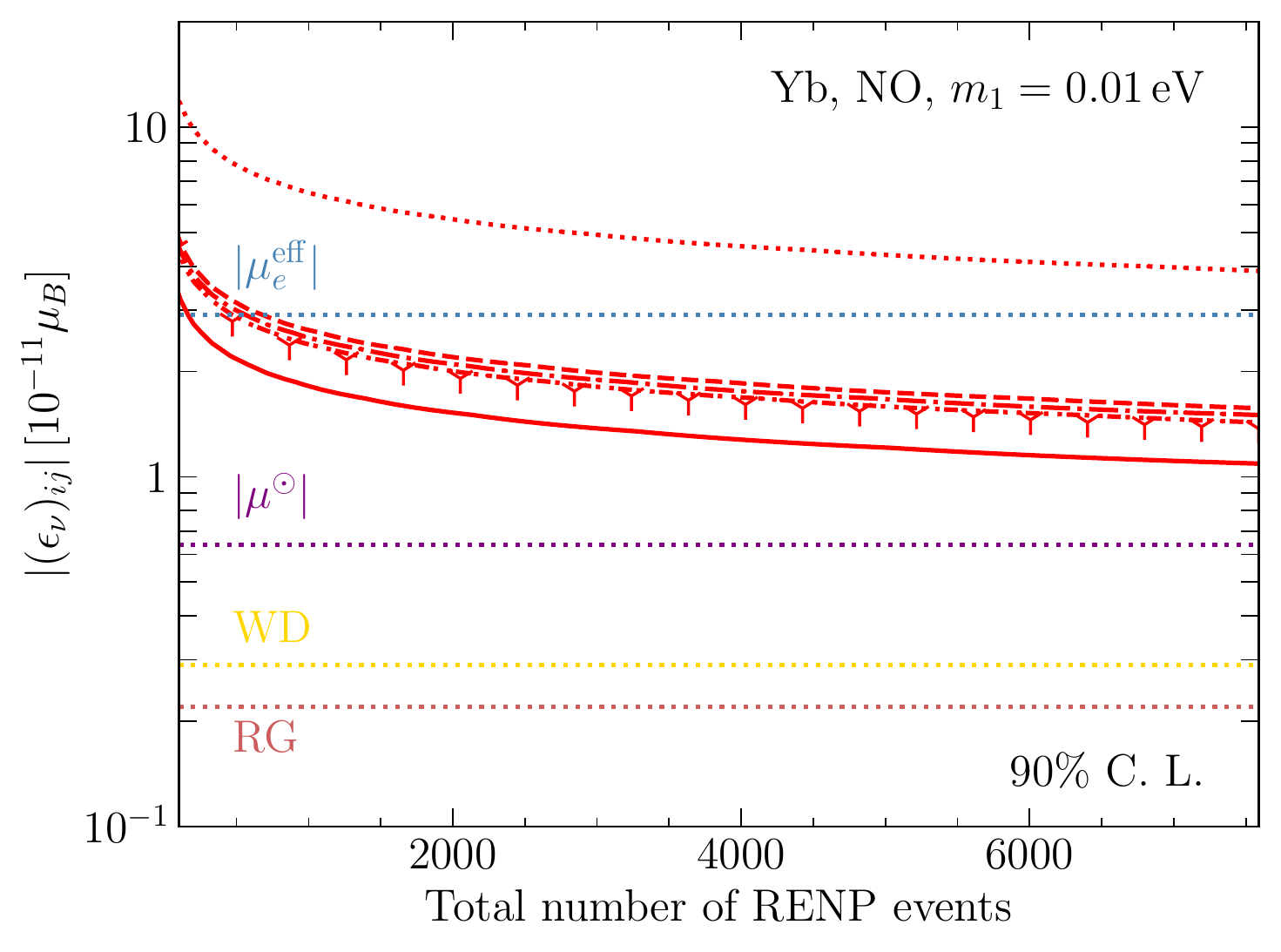}
\includegraphics[width=0.49\textwidth]{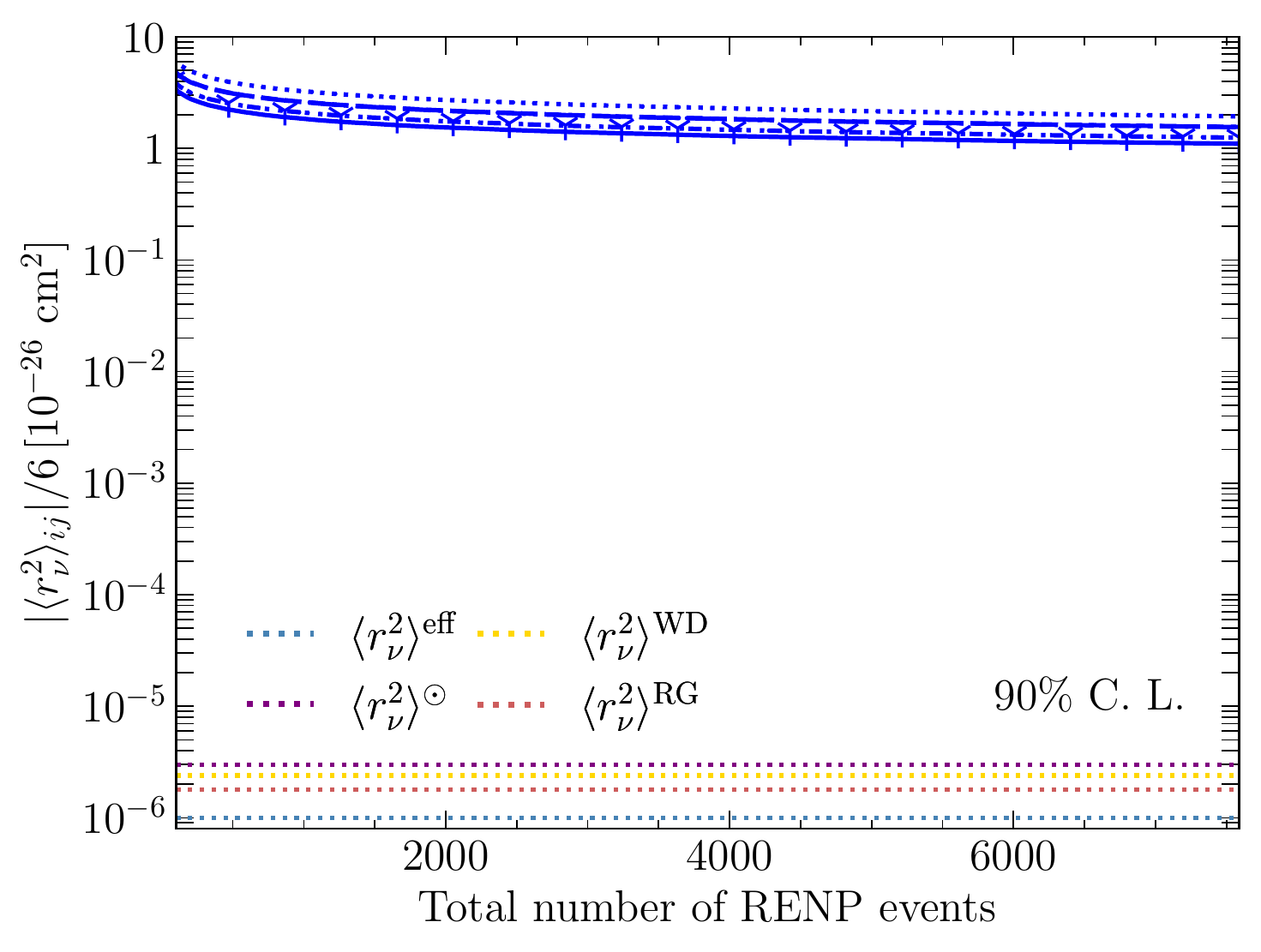}
\includegraphics[width=0.49\textwidth]{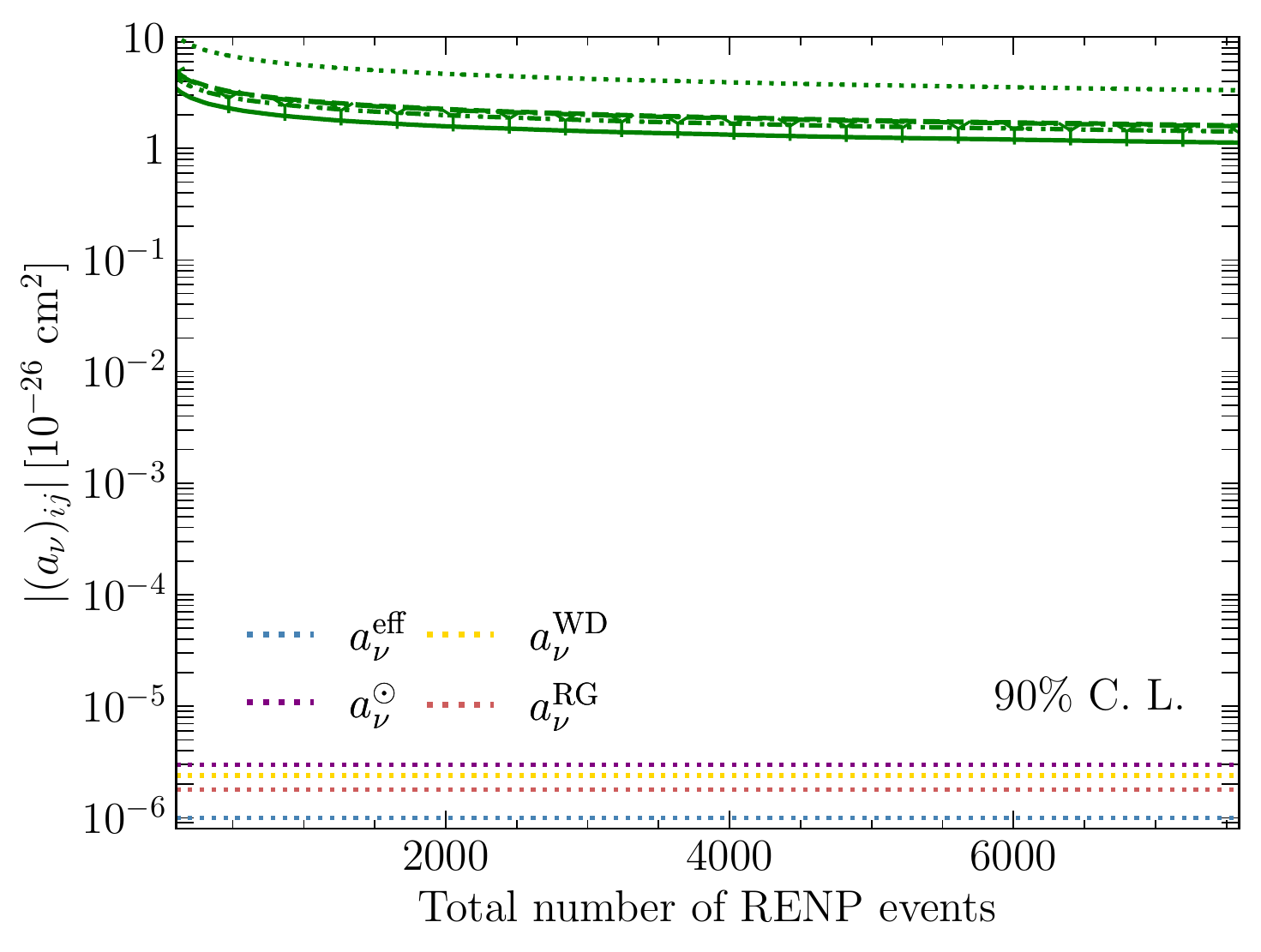}
\caption{The expected sensitivities to neutrino electromagnetic
form factors at 90\% C.L. as a function of the expected event
rate for $(\mu_\nu)_{ij}$ (top left),
$(\epsilon_\nu)_{ij}$ (top right), $\langle r_\nu^2\rangle_{ij}$
(bottom left), and $(a_\nu)_{ij}$ (bottom right) with the
normal ordering (NO) and $m_1 = 0.01$\,eV. The dotted lines
show the current bounds on the effective parameters from reactor
(dotted blue for $\mu_e^{\rm eff}$ or $\langle r_\nu^2\rangle^{\rm eff}$)
\cite{TEXONO:2009knm} and solar (dotted purple for $\mu_e^{\odot}$ or
$\langle r_\nu^2\rangle^{\odot}$) \cite{Khan:2017djo,
Borexino:2017fbd} neutrinos, as well as plasmon decay in red
giants (dotted red for RG) \cite{Diaz:2019kim} and
white dwarfs (dotted yellow for WD)
\cite{MillerBertolami:2014oki,Hansen:2015lqa}.
}
\label{fig:sensitivity}
\end{figure}

Following the setup in \cite{Ge:2022cib}, we estimate the 
number of RENP photons at each frequency $\omega$ as 
$N_i = T \Gamma_0 \mathcal I(\omega_i)$,
\begin{eqnarray}
   N_i
\approx 
  173 
  \left( \frac T {\rm days}\right)
  \left(\frac V {100~\rm cm^3}\right)
  \left( 
    \frac{n_a~{\rm or}~n_\gamma}{10^{21}\rm cm^{-3}}
  \right)^3
  \mathcal I(\omega_i),
\end{eqnarray}
where $T$ is the exposure time
that we take to be equal among all the nine $\omega_i$ 
values. While $V$ is the target volume, $\mathcal I(\omega)$
is the total spectral 
function containing the SM and EM contributions. 
For a typical Yb experiment with $T = 10$\,days, 
we expect $N(\omega_{i=1,\dots,9}) = (146$, 136, 120, 
107, 87, 82, 79, 39, 2.5) SM background events,
respectively, and in total 759 events.
The sensitivity is evaluated using the Poisson $\chi^2$
function,
\begin{eqnarray}
  \Delta \chi^2
\equiv
  2 \sum_{i = 1}^9
\Bigl\{
  N^{\rm true}_i - N^{\rm test}_i
- N^{\rm true}_i
  \log 
  \left[  N^{\rm true}_i/N^{\rm test}_i \right]
\Bigr\},
\label{eq:chi2_def}
\quad
\end{eqnarray}
where $N^{\rm true}_i$ ($N^{\rm test}_i$) are the 
event numbers with SM only (SM and EM) contributions at each 
frequency $\omega_i$. \gfig{fig:sensitivity} shows the 
90\% C.L. ($\Delta \chi^2 = 2.71$) sensitivity for
individual matrix elements as a 
function of the total number of events.
For comparison, we also show the existing
constraints from the reactor experiment TEXONO 
\cite{TEXONO:2009knm} (blue dotted line), the solar neutrino measurement 
\cite{Khan:2017djo,Borexino:2017fbd} (purple dotted line),
and the stellar cooling with plasmon decay for red giants (RG) 
\cite{Diaz:2019kim} (red dotted lines) or white dwarfs 
(WD) \cite{MillerBertolami:2014oki,Hansen:2015lqa} 
(yellow dotted lines) as discussed in 
\gsec{sec:current_bounds}. The sensitivity of $\mu_\nu$ 
reaches the direct detection experiment sensitivity of 
$3 \times 10^{-11}\mu_B$ at 1000 events while it takes
twice for $\epsilon_\nu$ to reach the same value.

As expected from the discussions in \gsec{sec:RENP_a_r}, 
the sensitivity for charge radius and anapole is 
around $10^{-26}$\,cm$^2$ which is 5 orders of magnitude 
smaller than the existing constraints. However, 
the existing constraints are on combinations of
various matrix elements and hence subject to parameter 
degeneracies. With much stronger constraints on
parameter combinations, the individual elements can
still be sizable. The only exception is stellar
cooling whose constraint is on the sum of matrix
elements squared. However, the stellar cooling
calculation suffers from large theoretical uncertainties.
The RENP experiment can provide unique capability
of identifying individual elements and their type.
We can also expect further sensitivity improvement
if the E1$\times$E1 configuration is adopted.
\begin{figure}[!t]
\centering
\includegraphics[width=0.48\textwidth]{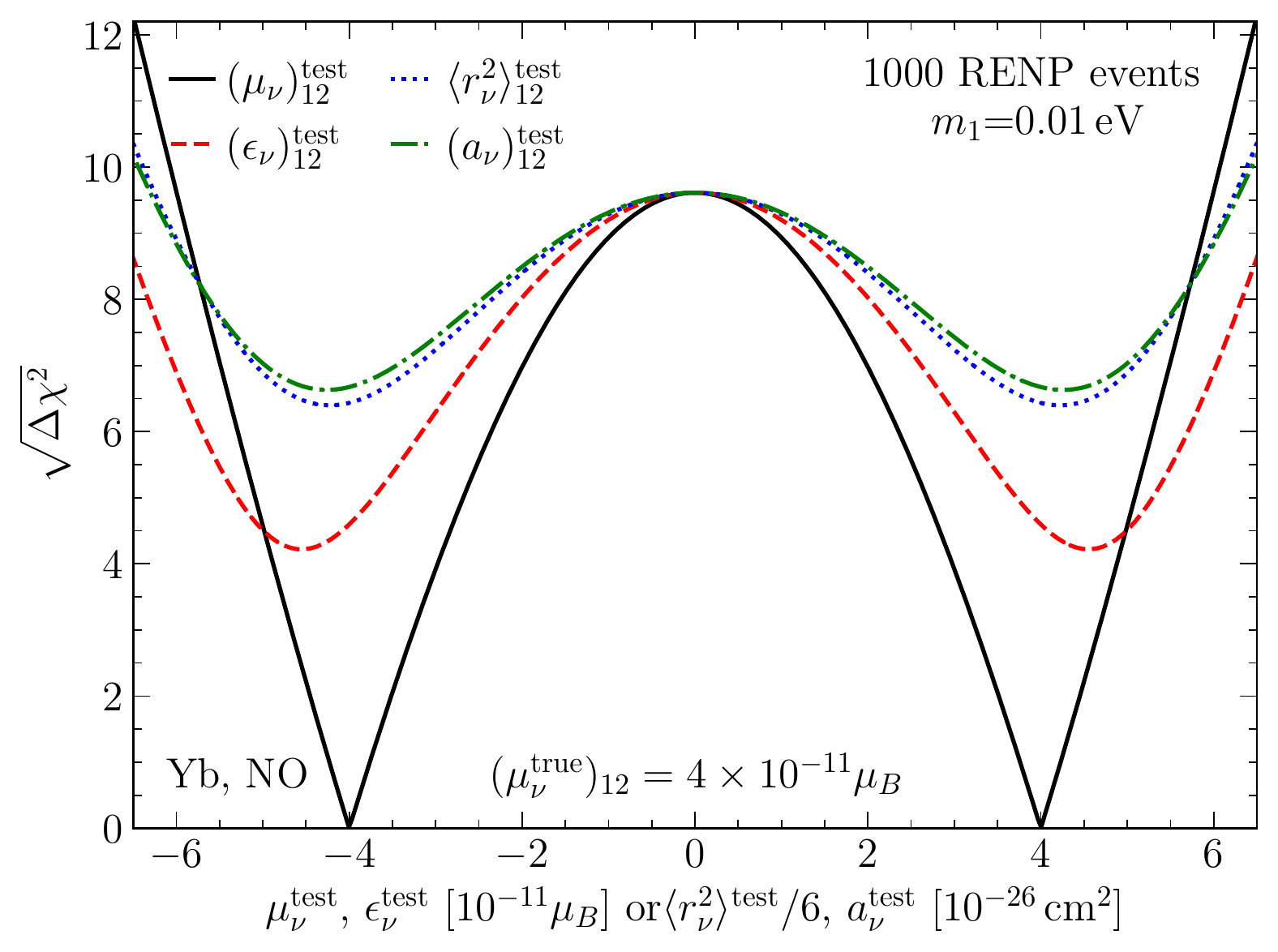}
\includegraphics[width=0.48\textwidth]{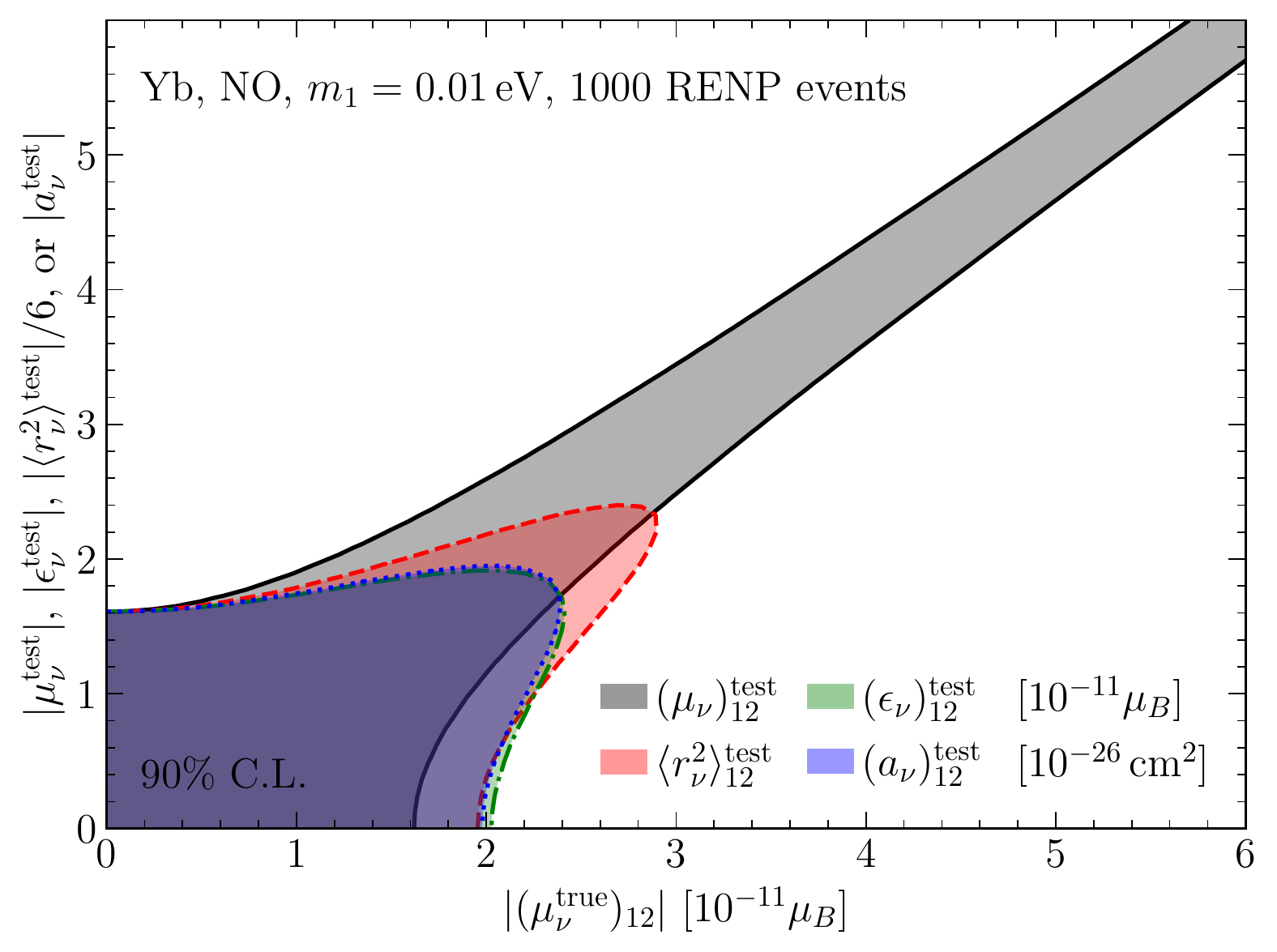}
\caption{{\bf Left :} The $\sqrt{\Delta\chi^2}$ profile
for a non-zero neutrino magnetic moment
$(\mu_\nu)_{12}^{\rm true} = 4 \times 10^{-11}\,\mu_B$. {\bf Right}: The allowed regions as a function of $(\mu_\nu)_{12}^{\rm true} \in [0,6]\times 10^{-11}\,\mu_B$. In total 1000
events are used for illustration. The regions represents several test models: $(\mu_\nu)_{12}^{\rm test}$ free (black), $(\epsilon_\nu)_{12}^{\rm test}$ free (red dashed), $\langle r_\nu^2\rangle_{12}^{\rm test} $ free (blue dotted), and $(a_\nu)_{12}^{\rm test}$ free (green dot-dashed).}
\label{fig:muSensitivity}
\end{figure}

To illustrate the RENP capability of distinguishing
the neutrino form factors, the left panel of 
\gfig{fig:muSensitivity} shows the $\Delta \chi^2$ for 
$(\mu_\nu)_{12}^{\rm true} = 4 \times 10^{-11}\,\mu_B$ when 
compared with the fit for $(\mu_\nu)_{12}^{\rm test}$, 
$(\epsilon_\nu)_{12}^{\rm test}$, $\langle r_\nu^2 
\rangle_{12}^{\rm test}$, or $(a_\nu)_{12}^{\rm test}$
with the same indices $i = 1$ and $j = 2$.
Since the decay rate is proportional to the 
coupling constant squared, all the curves have two local 
minimums with just a sign difference. While the black
curve with $(\mu_\nu)^{\rm test}_{12}$ as fitting variable
can touch the horizontal axis, the other
three curves have nonzero $\chi^2_{\rm min}$.
A non-zero $\Delta \chi^2$ minimum excludes the 
coupling by $\sqrt{\Delta \chi^2}$ standard deviation.
Both charge radius and anapole can be excluded by $\sim 
6.5\,\sigma$ while the neutrino EDM by
$4\,\sigma$ with 1000 RENP events ($T = 10$ days). 
The higher exclusion sensitivity for 
$\langle r_\nu^2\rangle_{12}$ and $(a_\nu)_{12}$ 
than the $(\epsilon_\nu)_{12}$ one is expected since
the shape of the EDM spectral function 
is closer to the magnetic moment one. The 
difference is especially significant at the lower 
frequency region that contains more events and
larger deviation as pointed above.

The right panel of \gfig{fig:muSensitivity} 
shows the sensitivity on the neutrino form factors
as a function of the true value $(\mu_\nu)_{12}^{\rm true}$
with the 90\% C.L. upper/lower limits on the 
test form factor elements that are shown as the $y$-axis.
For a nonzero $(\mu_\nu)_{12}^{\rm true}$, only the test 
magnetic moment has upper and lower bounds 
$(\mu_\nu)_{12}^{\rm test} \pm \Delta \mu_\nu$  
when $(\mu_\nu)_{12}^{\rm true} > 3 \times 
10^{-11}\mu_B$. Conversely, the EDM
parameter space cannot explain the data at 90\% C.L. 
for $(\mu_\nu)_{12}^{\rm true} > 3 \times 
10^{-11}\mu_B$ and has only an upper bound if 
$(\mu_\nu)_{12}^{\rm true} < 2\times 10^{-11}\mu_B$. 
Since the charge radius and anapole curves are 
significantly different from the magnetic moment one, 
they cannot explain the data at 90\% C.L. for 
$(\mu_\nu)_{12}^{\rm true} >2.4\times 10^{-11}\mu_B$.

\section{Discussion \& Conclusion}
\label{sec:Conclusion} 

The EM properties are not just the first neutrino interaction
that was originally proposed by Pauli in exactly the same
letter where he pointed out the existence of neutrino,
but can also have sizable values to appear in various BSM models.
Hence, observing the neutrino EM properties (magnetic
and electric moments, charge radius, and anapole) can serve
as a window of probing possible new physics. Unique
probe and identification of each form factor elements are
necessary which is a serious challenge for the existing
experimental measurements and observations. We show that 
the RENP process is an ideal probe. The typically eV scale
momentum transfer of atomic transitions is perfect
for probing light mediators including the massless photon.
Most importantly, the RENP process do not suffer from 
degeneracies in EM form factors and hence has the possibility
of individual identification as a unique feature. Both
analytical derivations and practical searching strategies
are provided with great details in the current paper.
In addition, the selection rules for the QED double photon
emission, the RENP with weak interactions, and the one with
neutrino EM properties are systematically derived. While
the RENP with weak interactions is dominated by the M1
transition, its counterpart with neutrino EM form factors
is actually similar to the QED photon emission that is
mainly contributed by the E1 transition. In other words,
the bounds can be probably further improved if a RENP experiment
can be done with E1$\times$E1 type atomic transitions
instead of the usual proposal with E1$\times$M1 transitions.

\section*{Acknowledgements}
The authors are supported by the National Natural Science
Foundation of China (12375101, 12090060, 12090064, and 12247141) and the SJTU Double First
Class start-up fund (WF220442604).
SFG is also an affiliate member of Kavli IPMU, University of Tokyo. PSP is also supported by the Grant-in-Aid for Innovative Areas No. 19H05810.

\appendix 

\section{Atomic Transitions with QED and EW Interactions}
\label{App:Htot}

\subsection{Non-Relativistic Electron Hamiltonian}
\label{App:EWHamiltonian}

Here we start from first principles and obtain the 
complete Hamiltonian for a non-relativistic bound
electron. It is more convenient to add the electron
QED and EW interactions in terms of Lagrangian,
\begin{eqnarray}
  \mathcal L_{EW}
=
  \overline e 
\left[
  i \slashed \partial 
- e \slashed A
- \sqrt{2} G_{\rm F} \slashed j^{(\nu)}
  \left( v + a \gamma_5 \right)
- m_e
\right] e
+ \dots, 
\end{eqnarray}
where $j^{(\nu)}_\mu \equiv \bar \nu \gamma_\mu P_L \nu$ is
the neutrino current. Note that the chiral coefficients
$v$ and $a$ are attached to the electron bilinears.
The Dirac equation for the electron field $e(x)$
in the presence of a photon field $A_\mu$ and a
neutrino current is,
\begin{eqnarray}
\left[ 
    i \partial_\mu
  + (J_V)_\mu
  - (J_A)_\mu \gamma_5
\right]
  \gamma^\mu e(x)
=
  m_e e(x).
  \label{eq:DiracEqbounded}
\end{eqnarray}
For convenience, we define the two 
effective vector and axial currents,
respectively, as
\begin{eqnarray}
  (J_V)_\mu
\equiv 
- e A_\mu 
- \sqrt{2} v G_{\rm F}  j^{(\nu)}_\mu,
\quad {\rm and} \quad 
  (J_A)_\mu
\equiv 
- \sqrt{2} a G_{\rm F}  j^{(\nu)}_\mu.
\label{eq:JVJAdef}
\end{eqnarray}

In the non-relativistic limit, the four-component
electron spinor would reduce to its two-component
counterpart that people are usually familiar with
in quantum mechanics while the other two components
are suppressed. This feature becomes more transparent
in the Dirac representation,
\begin{eqnarray}
  \gamma^0_D
=
\left( 
\begin{matrix}
  1 & 0 \\ 
  0 & -1 
\end{matrix}
\right),
\quad
  \gamma^i_D 
=
\left( 
\begin{matrix}
  0 & -\sigma^i  \\ 
\sigma^i  & 0  
\end{matrix}
\right),
\quad {\rm and} \quad 
  \gamma^5_D 
=
\left( 
\begin{matrix}
  0   & 1 \\ 
  1   & 0
\end{matrix}
\right).
\end{eqnarray}
Correspondingly, the electron field splits
into two two-component spinors, $e(x) \equiv (\psi({\boldsymbol 
x}), \phi({\boldsymbol x}))$ $e^{- i Et}$,
with $E$ being the total energy \cite{Sakurai:2011zz}.
Since our discussions in this paper focus on only
the electron particle and do not touch positron,
only the positive-frequency mode needs to be
considered for simplicity. While the time dependence
can be factorized out with $e^{- i E t}$, the
spatial distribution of the bound electron field in
an atom is nontrivial and will be elaborated below.
Putting the electron field $e(x)$ back into
\geqn{eq:DiracEqbounded} allows us to 
obtain two coupled equations for $\psi$ and $\phi$,
\begin{subequations}
\begin{eqnarray}
  \label{eq:psi_electron}
\left[
   (i \boldsymbol\nabla  + {\boldsymbol J}_V )
    \cdot \boldsymbol\sigma
  + J_A^0
 \right]
 \phi
=
- \left[ 
   E - m_e + J_V^0 + ({\boldsymbol J}_A \cdot \boldsymbol\sigma)
  \right]
  \psi 
  \,,
\\ 
 \left[
    (i  \boldsymbol \nabla +{\boldsymbol J}_V   )
    \cdot \boldsymbol\sigma
  + J^0_A
 \right]
 \psi
=
-\left[ 
     E + m_e + J_V^0 + ({\boldsymbol J}_A \cdot \boldsymbol\sigma)
  \right]
  \phi
  \,.
  \label{eq:phi_electron}
\end{eqnarray}
\label{eq:Electron_EoM}
\end{subequations}
Notice that the crucial signs of $E- m_e$ in the 
first equation and $E + m_e$ in the second one make
big difference. For a non-relativistic electron
($|{\boldsymbol p}_e|$, $J_V^0$, and ${\boldsymbol J}_A\cdot \boldsymbol \sigma \ll m_e$),
it follows from the second equation above that,
$\phi/\psi \sim  \mathcal O \left(|
{\boldsymbol p}_e|/m_e\right)$. Consequently, 
the electron wave function is given mostly by $\psi$
while the $\phi$ obtained from \geqn{eq:phi_electron},
\begin{eqnarray}
  \phi
\approx 
- \frac 1 {2m_e} 
\left[
  (i \boldsymbol \nabla  +{\boldsymbol J}_V)
  \cdot \boldsymbol\sigma
+J^0_A
\right]
  \psi,
\label{eq:Phi}
\end{eqnarray}
is highly suppressed.

For convenience, we redefine the total energy $E$
as $E \equiv m_e + E^{\rm NR}$, with
$E^{\rm NR} \ll m_e$. The quantity $E^{\rm NR}$ is the
non-relativistic energy of the electron. For a free 
electron, $E^{\rm NR} = T_e \approx |{\boldsymbol 
p}_e|^2/2m_e$. However, in the presence of 
interacting fields $E^{\rm NR}$ represents not
only the electron kinetic energy ($T_e$),
but contains potential terms related to $J_V$ and 
$J_A$, as will become evident below. Using the above 
approximation \geqn{eq:Phi} to substitute $\phi$
back into the left-hand side of \geqn{eq:psi_electron}
and keeping the leading order in $m_e^{-1}$, we obtain
the non-relativistic electron Schrödinger equation,
\begin{eqnarray}
  \left\{ 
\frac{1}{2m_e} 
 \left[
    (i \boldsymbol\nabla +{\boldsymbol J}_V)
    \cdot \boldsymbol\sigma
  + J_A^0
 \right]^2
-
  J_V^0
-
  ({\boldsymbol J}_A \cdot \boldsymbol\sigma)
  \right\} 
  \psi 
=
  E^{\rm NR} \psi
  \,.
\end{eqnarray}
In other words, the electron Hamiltonian with both
QED and EW interactions is,
\begin{eqnarray}
    H_{EW}
\equiv 
\frac{1}{2m_e} 
 \left[
    (i \boldsymbol\nabla + {\boldsymbol J}_V)
    \cdot \boldsymbol\sigma
   +J_A^0
 \right]^2
-
  J_V^0
-
  ({\boldsymbol J}_A \cdot \boldsymbol\sigma).
\label{eq:H_EW}
\end{eqnarray}
The first term is a kinetic energy in 
the presence of vector/pseudo-vector fields while the 
last two terms are the potential energy.

For later convenience, we reformulate the Hamiltonian as,
\begin{subequations}
\begin{align}
  H_{EW}
& =
  \frac 1 {2m_e} 
 \left[
     ( i \boldsymbol\nabla +{\boldsymbol J}_V)^2
   +2 J_A^0 \boldsymbol\sigma\cdot 
      (i \boldsymbol\nabla + {\boldsymbol J}_V )
    + i\boldsymbol\sigma \cdot (\boldsymbol \nabla  J_A^0)
    +
      (J_A^0)^2
    -  \boldsymbol \sigma \cdot (\boldsymbol \nabla \times {\boldsymbol J}_V) 
 \right]
\\
&
- J_V^0
- {\boldsymbol J}_A \cdot \boldsymbol\sigma,
\end{align}
\label{eq:H_EWJvJafinal}
\end{subequations}
by using the identity
$(\boldsymbol{\sigma}\cdot {\boldsymbol A} ) 
(\boldsymbol{\sigma}\cdot  {\boldsymbol B}) = {\boldsymbol A}\cdot 
{\boldsymbol B} + i \boldsymbol{\sigma} \cdot ({\boldsymbol A}\times 
{\boldsymbol B})$. The total Hamiltonian can split into
three parts,
\begin{eqnarray}
    H_{EW} 
\equiv
  H_0 + H_\gamma + H_{\nu \bar \nu},
\label{eq:H0gnu}
\end{eqnarray}
for describing the atomic energy levels, the atomic
transitions, and the radiative emission of neutrino
pair as we elaborate below.

\subsection{Atomic Energy Levels and Transition with QED}
\label{sec:Hemission}

The Hamiltonian in \geqn{eq:H_EWJvJafinal} can in principle
describe any vector and axial-vector interactions.
For electromagnetic coupling with photon,
$J_V^0 = -e A_0 = -e V(x)$ and ${\boldsymbol J}_V = -e 
{\boldsymbol A}$. To be exact, $J_V^0$ is the electric
potential due to the electron charge and
$\boldsymbol \nabla \times {\boldsymbol J}_V = - e \boldsymbol \nabla 
\times {\boldsymbol A}$ is actually the magnetic field
coupled with electron spin.

The photon field has two contributions,
$A_\mu \equiv A^{(0)}_\mu + A^{(\gamma)}_\mu$, a classical
EM field $A^{(0)}_\mu$ as background generated by the
atomic charges and a quantum fluctuation component
for photon emission/absorption $A^{(\gamma)}_\mu$.
Of these two components, the quantum part $A^{(\gamma)}_\mu$
is usually treated as radiative perturbations while
the classical one $A^{(0)}_\mu$ forms the atomic
Hamiltonian, $H_0$, that describes the atomic energy 
levels, $H_0 | a\rangle = E_a |a \rangle$ with $a = g$,
$e$, and $v$ for the ground, excited, and intermediate
virtual states. To be more concrete, $H_0$ is a
combination of the electron kinetic energy and
the classical electromagnetic potential obtained
by replacements $J_V \rightarrow - e A^{(0)}$
and $J_A \rightarrow 0$ from \geqn{eq:H_EWJvJafinal},
\begin{eqnarray}
  H_0 
\equiv 
- \frac{\boldsymbol\nabla^2}{2m_e}  
- \frac{ie}{2m_e}
\left(
  2 {\boldsymbol A}^{(0)} \cdot \boldsymbol\nabla
+ \boldsymbol\nabla\cdot {\boldsymbol A}^{(0)}
\right)
+ \frac{e^2}{2m_e} ({\boldsymbol A}^{(0)})^2
+ \frac{e\boldsymbol\sigma }{2m_e} \cdot (\boldsymbol\nabla \times {\boldsymbol A}^{(0)})
+ e A_0^{(0)}.
\quad
\label{eq:H0def}
\end{eqnarray}

Since $A^{(\gamma)}$ is a small perturbation,
one needs to keep only the linear terms of $A^{(\gamma)}$
for the photon emission Hamiltonian $H_\gamma$,
\begin{align}
  H_\gamma
& \equiv
 e A^{(\gamma)}_0
-  \frac{e}{m_e} {\boldsymbol A}^{(\gamma)} \cdot 
  (i\boldsymbol\nabla - e  {\boldsymbol A}^{(0)})
- \frac{ie}{2m_e}
  \boldsymbol \nabla \cdot {\boldsymbol A}^{(\gamma)}
+  \frac{e}{2m_e} \boldsymbol\sigma \cdot 
  (\boldsymbol\nabla \times {\boldsymbol A}^{(\gamma)}).
\end{align}
Since the single-photon emission
($|i \rangle \rightarrow |f\rangle + \gamma$)
happens between two different atomic states
($|i \rangle$ for the initial and $|f \rangle$ for
the final), the trivially multiplicative operators
$e A^{(\gamma)}_0$ and
$i e \boldsymbol \nabla \cdot {\boldsymbol A}^{(\gamma)} / 2 m_e$
cannot contribute due to orthogonality of atomic
states. For example, $\langle i | 
e A^{(\gamma)}_0 |f\rangle  = e A^{(\gamma)}_0 \langle i | 
f\rangle = 0$.
Although the electric potential $e A^{(\gamma)}_0$
contributes mainly for determining the atomic energy levels
but not the transition among them. So finally there are only
two terms remaining.

The term $i\boldsymbol\nabla - e  {\boldsymbol A}^{(0)}$ 
is related to the position operator $\hat{\boldsymbol  x}$ via
$i\boldsymbol\nabla - e  {\boldsymbol A}^{(0)} = i m_e [ \hat{\boldsymbol x}, H_0]$.
With this replacement, the photon emission Hamiltonian becomes,
\begin{align}
  H_\gamma
& = 
- i e {\boldsymbol A}^{(\gamma)} \cdot [\hat{\boldsymbol  x}, H_0]
+ \frac{e}{2m_e} \boldsymbol\sigma \cdot 
  (\boldsymbol\nabla \times {\boldsymbol A}^{(\gamma)}).
\label{eq:HgammaAPP}
\end{align}
The first is proportional to the space coordinate operator
$\hat{\boldsymbol  x}$ and hence contributes as dipole operator,
${\boldsymbol d}_{if} \propto \langle i | \hat {\boldsymbol x} |f\rangle$.
On the other hand, the second term contains the electron
spin operator ($\hat {\boldsymbol s} = \boldsymbol \sigma / 2$)
and the magnetic field
($\boldsymbol B = \boldsymbol \nabla \times A^{(\gamma)}$).
Both terms are contributed
by the three-dimensional vector potential $\boldsymbol A^{(\gamma)}$.

The two terms in \geqn{eq:HgammaAPP} have definite spin and
parity such that the induced transitions are subject to
handy selection rules for the single-photon emission,
$|i \rangle \rightarrow |f\rangle + \gamma$.
The total angular momentum from the initial ($J_i$) to the 
final one ($J_f$) changes by at most one single unity,
$J_f - J_i = 0$, $\pm 1$, with the $J_i = J_f = 0$
transition also forbidden. At the same time,
the magnetic quantum number should also change at most by one 
unity, $M_{J_i} - M_{J_f} = 0$, $\pm 1$. The relative parity between 
the initial ($\pi_e$) and final ($\pi_g$) states determines if the 
transition is E1 type with parity flip ($\pi_f = - \pi_i$) or
M1 type with parity conserved ($\pi_f = \pi_i$).
Since the dipole operator reverses sign under parity
transformation,
$\hat{\boldsymbol x} \rightarrow - \hat {\boldsymbol x}$, it
induces E1 type transition while the magnetic moment operator
in the second term of \geqn{eq:HgammaAPP} is M1 type.

The two terms in \geqn{eq:HgammaAPP} have different 
typical sizes. The dipole term
is in unit of the Bohr radius $a_0 = (m_e \alpha)^{-1}$
while the magnetic moment operator is regulated by the Bohr 
magneton ($\mu_B = e/2m_e$) which is suppressed by the 
electron mass. Although the dipole term also contains $H_0$
that gives the atomic energy at $\mathcal O($eV), the
$\boldsymbol \nabla$ operator in the magnetic moment operator
gives the photon frequency which is also at eV scale
to balance. Consequently, the E1 transition is
typically $\alpha^{-1}\sim 100$ times larger than
its M1 counterpart.

\subsection{Radiative Emission of Neutrino Pair with EW}
\label{sec:Hew}

The remaining terms in the Hamiltonian of \geqn{eq:H0gnu}
are responsible for the neutrino pair emission,
$|e \rangle \rightarrow |v\rangle + \nu \bar \nu$,
\begin{subequations}
\begin{align}
 H_{\nu\bar\nu} 
& = 
- i \sqrt{2} G_{\rm F} 
  \left( 
     v{\boldsymbol j}^{(\nu)}   
   + a j_0^{(\nu)} \boldsymbol \sigma 
  \right)
  \cdot [\hat{\boldsymbol  x}, H_0]
+
 \sqrt 2 G_{\rm F}
 (v j^{(\nu)}_0 + a {\boldsymbol j}^{(\nu)} \cdot \boldsymbol \sigma)
\label{eq:Hnunua}
\\ & 
+\frac{v G_{\rm F} }{\sqrt{2} m_e}
  \boldsymbol \sigma\cdot(\boldsymbol\nabla \times {\boldsymbol j}^{(\nu)} )
-  \frac{i G_{\rm F} }{\sqrt{2}m_e}
  \boldsymbol \nabla \cdot 
  \left( 
     v {\boldsymbol j}^{(\nu)} 
   + a j_0^{(\nu)}\boldsymbol \sigma 
  \right).
\label{eq:Hnunub}
\end{align}    
\end{subequations}
While those terms in the second line are suppressed by
the electron mass, any meaningful contribution arises
from those in the first line. Due to orthogonality,
the $\sqrt 2 G_{\rm F} v j^{(\nu)}_0$ term does not
contribute to the atomic transition. Among the remaining
three terms, $v {\boldsymbol j}^{(\nu)} \cdot [\hat {\boldsymbol x}, H_0]$
with dipole operator is E1 type while
$a {\boldsymbol j}^{(\nu)} \cdot \boldsymbol \sigma$ with spinor
operator is M1 type. Contrary to the QED emissions discussed
above, the neutrino pair M1 emission is not suppressed
by the electron mass and hence is around 3 orders
($\sim E_{ve}/\alpha m_e$ as a product of the atomic energies
and the Bohr radius)
of magnitude larger than its E1 counterpart. In other words,
the leading photon emission is E1 type while the leading
neutrino pair emission is M1 type. Finally, the
$a j^{(\nu)}_0 \boldsymbol \sigma \cdot [\hat {\boldsymbol x}, H_0]$
term is a product of the magnetic moment and dipole
operators which is at most of the same size
as the E1 type and hence subleading.

\end{document}